\documentclass{aa}

\usepackage{txfonts}
\usepackage{graphicx}
\usepackage{natbib}
\bibpunct{(}{)}{;}{a}{}{,}

\begin{document}

\title{Spectro-photometric close pairs in GOODS-S: major and minor companions of intermediate-mass galaxies}
\titlerunning{Spectro-photometric major and minor companions in GOODS-S}
\authorrunning{C. L\'opez-Sanjuan, et al.}

\author{Carlos L\'opez-Sanjuan\inst{1,2,3,4} \and Marc Balcells\inst{1,2,5} \and Pablo G. P\'erez-Gonz\'alez\inst{3,6} \and Guillermo Barro\inst{3} \and Jes\'us Gallego\inst{3} \and Jaime Zamorano\inst{3}}

\institute{Instituto de Astrof\'{\i}sica de Canarias, Calle V\'{\i}a L\'actea s/n, E-38205 La Laguna, Tenerife, Spain \and Departamento de Astrof\'\i sica, Universidad de La Laguna, E-38200 La Laguna, Tenerife, Spain \and Departamento de Astrof\'{\i}sica y Ciencias de la Atm\'osfera, Facultad de C.C. F\'{\i}sicas, Universidad Complutense de Madrid, E-28040 Madrid, Spain \and Laboratoire d'Astrophysique de Marseille, P\^ole de l'Etoile Site de Ch\^ateau-Gombert 38, rue Fr\'ed\'eric Joliot-Curie, F-13388 Marseille, France. \email{carlos.lopez@oamp.fr} \and Isaac Newton Group of Telescopes, Aptdo.~Correos 321, E-38700 Santa Cruz de La Palma, Tenerife, Spain \and Steward Observatory, University of Arizona, 933 North Cherry Avenue, Tucson, AZ 85721, USA}

\date{}

\abstract
{}{Recent work has shown that major mergers of disc galaxies can only account for $\sim$20\% of the growth of the galaxy red sequence between $z=1$ and $z=0$.  
Our goal here is to provide merger frequencies that encompass both major and minor mergers,  derived from close pair statistics. We aim to show that reliable close pair statistics can be derived from galaxy catalogues with mixed spectroscopic and photometric redshifts.}
{We use $B-$band luminosity- and mass-limited samples from a {\it Spitzer}/IRAC-selected catalogue of GOODS-S.  We present a new methodology for computing the number of close companions, $N_\mathrm{c}$, when spectroscopic redshift information is partial. The methodology extends the one used in spectroscopic surveys to make use of photometric redshift information. We select as close companions those galaxies separated by $6h^{-1}$ kpc $< r_{\rm p} < 21h^{-1}$ kpc in the sky plane and with a difference $\Delta v \leq 500$ km s$^{-1}$ in redshift space.}
{We provide $N_\mathrm{c}$ for four different $B-$band-selected samples. It increases with luminosity, in good agreement with previous estimations from spectroscopic surveys.  
The evolution of $N_\mathrm{c}$ with redshift is faster in more luminous samples. We provide $N_\mathrm{c}$ of $M_{\star} \geq 10^{10}\ M_{\odot}$ galaxies, finding that the number including minor companions ($N_{\rm c}^{\rm m}$, mass ratio $\mu \geq 1/10$) is roughly two times the number of major companions alone ($N_{\rm c}^{\rm M}$, mass ratio $\mu \geq 1/3$) in the range $0.2 \leq z < 1.1$. We compare the major merger rate derived by close pairs with the one computed by morphological criteria, finding that both approaches provide similar merger rates for field galaxies when the progenitor bias is taken into account. Finally, we estimate that the total (major+minor) merger rate is $\sim$1.7 times the major merger rate.}
{Only 30\% to 50\% of the $M_{\star} \geq 10^{10}\ M_{\odot}$ early-type (E/S0/Sa) galaxies that appear between $z=1$ and $z=0$ may have undergone a major or a minor merger. Half of the red sequence growth since $z=1$ is therefore unrelated to mergers.}


\maketitle

\section{Introduction}\label{intro}
How important are galaxy mergers on the mass assembly history of red sequence (i.e., passive early-type) galaxies? This is one of the most challenging questions in galaxy evolution studies, one that is motivated by both theory and observations. In the former, popular hierarchical $\Lambda$-CDM models predict that the more massive dark matter haloes, inhabited by red-sequence galaxies, are the final stage of successive mergers of less massive haloes. However, the behaviour of the baryonic component is still unclear. Only with many ad-hoc ingredients can the latest models, which include radiative cooling, star formation, and AGN and supernova feedback, reproduce the observational trends better \citep[see][and references therein]{bower06,delucia07,stewart09,hopkins09bulges}. On the other hand, $N$-body simulations suggest that gas-rich mergers can produce intermediate-mass spheroidal systems \citep[e.g.,][]{naab03,bournaud05,hopkins08ss}, while dissipationless mergers can explain the more massive spheroids \citep[e.g.,][]{gongar03,gongar05,naab06ee}. Observationally, merger remnants in the local Universe can evolve into elliptical galaxies \citep{rothberg06a,rothberg06b}. In addition, the size \citep{daddi05,trujillo07,buitrago08,vanderwel08esize} and velocity-dispersion \citep{cenarro09} evolution of massive galaxies with redshift, and the luminosity density evolution of red sequence galaxies since $z \sim 1$ \citep{bell04,faber07} rule out the passive-evolution hypothesis and suggest galaxy mergers are an important process in galaxy formation and growth.  

The merger fraction, $f_{\rm m}$, defined as the ratio between the number of merger events in a
sample and the total number of sources in the same sample, is a useful observational quantity to explore the role of mergers in the growth of the red sequence and thus to constraint cosmological models. Many 
studies have determined the merger fraction and its evolution with redshift up to $z \sim 1$, 
usually parametrized as $f_{\rm m}(z) = f_{\rm m}(0) (1+z)^m$, using different
sample selections and methods, such as morphological criteria \citep[]{conselice03ff,conselice08,conselice09cos,lavery04,cassata05,lotz08ff,bridge07,kamp07,jogee09,clsj09ffgs,heiderman09,darg10i},
 kinematic close companions \citep[]{patton00,patton02,patton08,lin04,lin08, 
depropris05,depropris07}, spatially close pairs \citep[]{lefevre00,bundy04,bundy09,bridge07,kar07,hsieh08,bluck09}, or the correlation function \citep[]{bell06, 
masjedi06}. In these studies the value of the merger index $m$ at redshift $z 
\lesssim 1$ varies in the range $m =$ 0--4. $\Lambda$-CDM models predict $m 
\sim$ 2--3 \citep[]{kolatt99, governato99, gottlober01,fak08} for dark matter 
haloes, while suggesting a weaker evolution, $m \sim$ 0--2, for the galaxy 
merger fraction \citep{berrier06,stewart08}.

In this paper we study the number of close companions ($N_{\rm c} \sim 2f_{\rm m}$) in a {\it Spitzer}/IRAC-selected catalogue of the GOODS-S area and explore its dependence on redshift, $B$-band luminosity, and stellar mass. A robust methodology for measuring $N_{\rm c}$ in spectroscopic surveys was developed by \citet{patton00}, and we adapt it in present paper to exploit all the available redshift information in spectro-photometric catalogues, obtaining reliable values when compared with those from fully spectroscopic samples. In addition, work with close companions gives us useful information about the galaxies involved in the mergers, such as their mass ratio. Thanks to that, and taking advantage of our new methodology, we report an estimation of the number of minor companions (stellar mass ratio 1:10) and their relation with major companions (mass ratio higher than 1:3) for intermediate mass galaxies \citep[see][for a morphological determination of the minor merger fraction]{lotz08ff,jogee09}.

The paper is organized as follows. In Sect.~\ref{data} we summarize the GOODS-S data set we used, and in Sect.~\ref{metodo} we develop the methodology for determining the number of companions in spectro-photometric catalogues. Then, in Sect.~\ref{ncb} we study the number of close companions in $B$-band selected samples, while mass-selected samples are analysed in Sect.~\ref{ncmass}. We compare our inferred major merger rates with those from morphological criteria in Sect.~\ref{mrpair}. We discuss the implications of our results in Sect.~\ref{discussion}, and in Sect.~\ref{conclusion} we present our conclusions. We use $H_0 = 70\ {\rm km\
s^{-1}\ Mpc^{-1}}$, $\Omega_{M} = 0.3$, and $\Omega_{\Lambda} = 0.7$ throughout.
All magnitudes are Vega unless otherwise noted.

\section{GOODS-S catalogue}\label{data}
We worked with the galaxy catalogue from 
the Great Observatories Origins Deep Survey South (GOODS-S)\footnote{http://www.stsci.edu/science/goods/}  field by the {\it Spitzer} Legacy Team
\citep{giavalisco04}. We used the Version 1.0
catalogues\footnote{http://archive.stsci.edu/prepds/goods/} and reduced mosaics in
the $F435W$ ($B_{435}$), $F606W$ ($V_{606}$), $F775W$ ($i_{775}$), and $F850LP$
($z_{850}$) {\it HST}/ACS bands. These catalogues were cross-correlated using a
$1.5^{\prime\prime}$ search radius with the GOODS-S IRAC selected sample in the
Rainbow cosmological database\footnote{http://guaix.fis.ucm.es/$\sim$pgperez/Proyectos/ucmcsdatabase.en.html} published in \citet[see also \citealt{pgon05} and Barro et al., in prep.]{pgon08}, which provided us with spectral energy
distributions (SEDs) in the UV-to-MIR range, well-calibrated and with reliable
photometric redshifts, stellar masses, star formation rates, and rest-frame
absolute magnitudes. We worked with a {\it Spitzer}/IRAC selected catalogue in $[3.4]$ and $[4.5]$ filters to ensure completeness in stellar mass. Although the PSF of {\it Spitzer}/IRAC is $\sim2^{\prime\prime}$ \citep{fazio04}, we were able of resolve sources with $\sim1^{\prime\prime}$ separation (see \citealt{pgon08} for details about the deblending process of {\it Spitzer}/IRAC sources). We used this $1^{\prime\prime}$ separation to fix the minimum radius when searching for close companions (Sect.~\ref{ncs}).

We refer the reader to these papers for a more detailed description of
the data included in the SEDs and the analysis procedure. Here, we briefly summarize the main characteristics of the data set. The Rainbow database contains consistent
aperture photometry in several UV, optical, NIR, and MIR bands with the method
described in \citet{pgon08}. The UV-to-MIR SEDs were built for $4927$ IRAC sources in the GOODS-S region down to a 75\% completeness magnitude $[3.6]$$=$23.5~mag
(AB). These SEDs were fitted to stellar population and dust emission models to
obtain estimates of the photometric redshift ($z_{\rm phot}$), the stellar
mass ($M_{\star}$), and the rest-frame $B$-band absolute magnitude ($M_B$).

Rest-frame absolute B-band magnitudes were
estimated for each source by convolving the templates fitting the SED with the
transmission curve of a typical Bessel-$B$ filter, taking the
redshift of each source into account. This procedure provided accurate interpolated
$B$-band magnitudes including a robustly estimated $k$-correction. Stellar
masses were estimated using the exponential star formation PEGASE01 models with
a \citet{salpeter55} IMF, and various ages, metallicities, and dust contents \citep{calzetti00} were included. The typical uncertainties in the stellar masses are a factor of $\sim$2, which is 
typical of most stellar population studies \cite[see, e.g.,][]{papovich06,fontana06}.

In the catalogue, $\sim40$\% of the sources have spectroscopic redshift ($z_{\rm spec}$), and we rely on $z_{\rm phot}$ for the other $\sim60$\%. Because of this, we refer to our catalogue as {\it spectro-photometric} hereafter. The median accuracy of the photometric redshifts at $z < 1.5$ is $|z_{\rm spec}
- z_{\rm phot}|/(1+z_{\rm spec}) = 0.04$, with a fraction $<$5\% of catastrophic
outliers \citep[][Fig.~B2]{pgon08}. In the present paper we use $\sigma_{z_{\rm phot}} =
\sigma_{\delta_z} (1+z_{\rm phot})$ as $z_{\rm phot}$ error, where
$\sigma_{\delta_z}$ is the standard deviation in the distribution of the
variable $\delta_z \equiv (z_{\rm phot} - z_{\rm spec}) / ({1 + z_{\rm phot}})$,
which is described by a Gaussian well with mean $\mu_{\delta_z} \sim 0$ and
standard deviation $\sigma_{\delta_z}$ (see \citealt{clsj09ffgs}, for details).
We take $\sigma_{\delta_z} = 0.043$ for $z \leq 0.9$ sources and $\sigma_{\delta_z} = 0.05$ for $z > 0.9$ sources.

Finally, we remove those sources in the catalogue within $\Delta z = 0.01$ of the centre of the most prominent large-scale structure (LSS) in GOODS-S, located at $z = 0.735$ \citep{ravikumar07,rawat08,clsj10megoods}. This is because the relative velocity of two galaxies located in a cluster is representative not of the dynamical state of the pair, but of the cluster potential, and we cannot apply the Sect.~\ref{metodo} methodology. For that reason, the present results are mainly refer to field galaxies.

\section{Methodology}\label{metodo}
In this section we recall the methodology developed by \citet{patton00}, which has been used extensively on spectroscopic samples \citep{patton00,patton02,patton08,lin04,lin08,depropris05,depropris07,depropris10,deravel09}. Then, we extend that methodology to use all the available information in spectro-photometric samples, paying attention to different bias as luminosity/mass (Sect.~\ref{ncs}) and spectroscopic (Sect.~\ref{fspec}) completeness of the samples, the border effects in redshift space and images limits (Sect.~\ref{border}), and the treatment of multiple systems (Sect.~\ref{ncrp}).

\subsection{Close pair statistics in spectroscopic samples}\label{ncs}
The linear distance between two sources can be obtained from their projected separation, $r_{\rm p} = \theta d_A(z_i)$, and their rest-frame relative velocity along the line of sight, $\Delta v = {c\, |z_l - z_i|}/(1+z_i)$, where $z_i$ and $z_l$ are the redshift of the primary (more luminous/massive galaxy in the pair) and secondary galaxy, respectively; $\theta$ is the angular separation, in arcsec, of the two galaxies on the sky plane; and $d_A(z)$ is the angular scale, in kpc/arcsec, at redshift $z$. Two galaxies are defined as a close pair if $r_{\rm p}^{\rm min} < r_{\rm p} \leq r_{\rm p}^{\rm max}$ and $\Delta v \leq \Delta v^{\rm max}$. The lower limit in $r_{\rm p}$ is imposed to avoid identifying bright star-forming regions of the primary galaxy as close companions. Common limits are $r_{\rm p}^{\rm min} = 5h^{-1}$ kpc, $r_{\rm p}^{\rm max} = 20h^{-1}$ kpc, and $\Delta v^{\rm max} = 500$ km s$^{-1}$. With these constraints 50\%-70\% of the selected close pairs will finally merge \citep{patton00,patton08,lin04,bell06}. We used the same limit in velocity but slightly different $r_{\rm p}$ limits: $r_{\rm p}^{\rm min} = 6h^{-1}$ kpc and $r_{\rm p}^{\rm max} = 21h^{-1}$ kpc. First, the minimum distance for which we are able to resolve two separate sources in our catalogue is $1\arcsec$ (Sect.~\ref{data}), which corresponds to $\sim 8.5$ kpc (6$h^{-1}$ kpc) in our cosmology at the minimum of the function $d_A(z)$. This confusion limit fixes the value of $r_{\rm p}^{\rm min}$. Second, we imposed $r_{\rm p}^{\rm max} - r_{\rm p}^{\rm min} = 15h^{-1}$ kpc. This condition comes from the study of \citet{bell06}. They find that the merger fraction is proportional to the radial range under study, so we kept the $15h^{-1}$ kpc range used widely in the literature.

To compute close pairs we defined a primary and a secondary sample. The primary sample contains the more luminous source of the pair, and we looked for those galaxies in the secondary sample that fulfil the close pair criterion for each galaxy of the primary sample.

If we work with $B$-band luminosity-selected samples, the primary sample comprises the galaxies in the catalogue with $M_{B,{\rm up}} < M_B \leq M_{B,1}^{\rm sel}$, while the secondary comprises $M_{B,{\rm up}} < M_B \leq M_{B,2}^{\rm sel}$ galaxies, where $M_{B,{\rm up}}$ is an upper limit in luminosity to avoid the different clustering properties of the most luminous galaxies \citep{patton00}. In every case, $M_{B,2}^{\rm sel} \geq M_{B,1}^{\rm sel}$. If we work with mass-selected samples, primary sample comprises $M_{\star,1}^{\rm sel} \geq M_{\star}$ sources, and secondary comprises $M_{\star,2}^{\rm sel} \geq M_{\star}$ sources. In this case, $M_{\star,2}^{\rm sel} \leq  M_{\star,1}^{\rm sel}$.
With the previous definitions the number of companions ($N_\mathrm{c}$) per primary galaxy is
\begin{equation}
N_{\rm c} = \frac{1}{N_{1}}\sum^{N_{1}}_{i} N_{\rm c}^i,\label{ncspec}
\end{equation}
where $N_1$ is the number of sources in the primary sample, and $N_{\rm c}^i$ the number of galaxies of the secondary sample that fulfil the close pair criterion for the primary galaxy $i$. Equation~(\ref{ncspec}) is valid for volume-limited samples, but we work with luminosity/mass-limited samples. To avoid incompleteness effects, \citet{patton00} define the function
\begin{equation}
S_N(z) = \frac{\int^{M_{B,{\rm lim}}(z)}_{M_{B,{\rm up}}} \Phi(M_{B},z) {\rm d}M_B}{\int^{M_{B}^{\rm sel}}_{M_{B,{\rm up}}} \Phi(M_{B},z) {\rm d}M_B},\label{snz}
\end{equation}
where $M_{B,{\rm lim}}(z)$ is the limiting magnitude of the catalogue at redshift $z$, $M_B^{\rm sel} = M_{B,1}^{\rm sel} [M_{B,2}^{\rm sel}]$ is the selection magnitude of the primary [secondary] sample, and $\Phi(M_{B},z)$ is the luminosity function in the $B$-band at redshift $z$. The definition of the function $S_N(z)$ for a mass-limited sample is similar \citep[see][for details]{ryan08}, and we take $S_N(z) = 1$ when $M_{B}^{\rm sel} \leq M_{B,{\rm lim}}(z)$, or $M_{\star}^{\rm sel} \geq M_{\star,{\rm lim}}(z)$. The limiting $M_B$ magnitude was determined in \citet{clsj09ffgoods}:
\begin{equation}
M_{B,{\rm lim}}(z) = -13.78 -12.66z + 11.18z^2 - 3.74z^3,
\end{equation}
defined as the third quartile in $M_B$ distribution at each redshift \citep{pgon08}. With this definition, the catalogue is complete for galaxies brighter than $M_B = -19.5$ up to z $\sim 1.3$. The limiting mass was defined in \citet{pgon08} as the 75\% completeness of the catalogue for passively evolving galaxies, and can be parametrized in the range of interest as
\begin{equation}
M_{\star,{\rm lim}}(z) = 9.47z^{1/8}\ M_{\odot}.
\end{equation}
With this definition the catalogue is complete for galaxies more massive than $M_{\star} = 6 \times 10^{9}\ M_{\odot}$ up to $z \sim 1.3$.

We parametrize the luminosity/mass function in Eq.~(\ref{snz}) with a Schechter function:
\begin{eqnarray}
\Phi(M,z) = {\rm x_1}\ln(10)\phi^{*}(z) \nonumber\\
\times [10^{\, {\rm x_2}(M^{*}(z)-M)}]^{1+\alpha (z)}\exp[-10^{\, {\rm x_2}(M^{*}(z)-M)}],\label{lfunc}
\end{eqnarray}
where $M = M_{B}$ or $M_{\star}$, $\log_{10}(\phi^{*}(z)/{\rm Mpc}^{-3}) = \phi_0 + \gamma(z - 0.5)$, $M^{*}(z) = M^{*}_0 + \delta(z - 0.5)$, $\alpha(z) = \alpha_0 + \psi(z-0.5)$, and x$_{1}$ and x$_{2}$ are constants to obtain the right normalization of the luminosity/mass function. We obtain the $B$-band luminosity function parameters from \citet{faber07}, while mass function parameters from \citet{pgon08}. For clarity we summarize all these parameters in Table~\ref{lftab}.

\begin{table*}
\caption{$B$-band luminosity and mass function parameters}
\label{lftab}
\begin{center}
\begin{tabular}{lcccccccc}
\hline\hline
Sample selection & x$_{1}$ & x$_{2}$ & $M^{*}_0$ & $\delta$ & $\phi_0$ & $\gamma$ & $\alpha_0$ & $\psi$\\
\hline
$M_B$		& 0.4 & 0.4  & -21.07 & -1.23 & -2.46 & -0.12 & -1.30 & 0\\
$M_{\star}$	&   1 & -1 &  11.23 &  0.13 & -2.72 & -0.56 & -1.22 & -0.041\\
\hline
\end{tabular}
\end{center}
\begin{footnotetext}
TNOTE. See Eq.~(\ref{lfunc}) for details about the meaning of each parameter.
\end{footnotetext}
\end{table*}

Finally, and following \citet{patton00}, the number of companions normalized to a volume-limited sample is
\begin{equation}
N_{\rm c} = \frac{\sum^{N_1}_{i} \big[S_N(z_i) \sum_{l}S_N(z_l)^{-1} \big]}{\sum^{N_{1}}_{i} S_N(z_i)},\label{ncspecvol}
\end{equation}
where the index $l$ covers all the close companions of the primary galaxy $i$.

\subsection{Close pair statistics in spectro -- photometric samples}\label{ncp}
The main problem in close pair studies with photometric samples is to constrain the redshift space condition. For example, the $\Delta v \leq 500$ km s$^{-1}$ condition at $z_i \sim 0.7$ implies $|z_l - z_i| \equiv \Delta z \leq 0.0045$. This condition is $\sim 15$ times less than the typical $z_{\rm phot}$ error at that redshift in our catalogue, $\sigma_{z_i} \sim 0.07$. When one or both galaxies in a close spatial pair have photometric redshift, we therefore cannot apply the methodology in Sect.~\ref{ncs}. To date a few works have used photometric catalogues to determine pair statistics: \citet{kar07} and \cite{bundy09} tackle the problem using a projection correction calculated in random samples on the plane of the sky, but keeping the redshift information of the sources. \cite{hsieh08} redefine the velocity criterion to $\Delta z = 2.5\sigma_{z_i}$ and apply a conventional projection correction. Finally, \citet{ryan08} also redefine the redshift criterion to $\Delta z = 2\sigma_{z_i}$, but no projection correction is applied. In this paper we present a new approximation to determine close pairs in spectro-photometric samples based on the methodology of \citet{patton00} and other previous photometric works.

We use the following procedure to define a close pair system: first we search for close spatial companions of a primary galaxy, with redshift $z_1$ and uncertainty $\sigma_{z_1}$, assuming that the galaxy is located at $z_1 - 2\sigma_{z_1}$. This defines the maximum $\theta$ possible for a given $r_{\rm p}^{\rm max}$ in the first instance. If we find a secondary galaxy with redshift $z_2$ and uncertainty $\sigma_{z_2}$ in the range $r_{\rm p} \leq r_{\rm p}^{\rm max}$ and with a given luminosity/mass with respect to the primary galaxy, then we study both galaxies in redshift space. For convenience, we assume below that every primary galaxy has, at most, one close companion in the secondary sample. In this case, our two galaxies could be a close pair in the redshift range
\begin{equation}
[z^{-},z^{+}] = [z_1 - 2\sigma_{z_1}, z_1 + 2\sigma_{z_1}] \cap [z_2- 2\sigma_{z_2}, z_2 + 2\sigma_{z_2}].
\end{equation}
Because of variation in the range $[z^{-},z^{+}]$ of the function $d_A(z)$, a sky pair at $z_1 - 2\sigma_{z_1}$ might not be a pair at $z_1 + 2\sigma_{z_1}$. We thus impose the condition $r_{\rm p}^{\rm min} < r_{\rm p} \leq r_{\rm p}^{\rm max}$ at all $z \in [z^{-},z^{+}]$, and redefine this redshift interval if the sky pair condition is not satisfied at every redshift. After this, our two galaxies define the close pair system $j$ in the redshift interval $[z^{-}_j,z^{+}_j]$, where the index $j$ covers all the close pair systems in the sample.

The next step is to define the number of companions associated at each close pair system $j$. For this, we suppose in the following that a galaxy $i$ in whatever sample is described in redshift space by a probability distribution $P_i\, (z_i\, |\, \eta_i)$, where $z_i$ is the source's redshift and $\eta_i$ are the parameters that define the distribution. If the source $i$ has a photometric redshift, we assume that

\begin{eqnarray}
P_i\, (z_i\, |\, \eta_i) = P_G\, (z_i\, |\, z_{{\rm phot},i},\sigma_{z_{{\rm phot},i}}) = \nonumber\\
\frac{1}{\sqrt{2\pi}\sigma_{z_{{\rm phot},i}}}\exp\left\{{-\frac{(z_i-z_{{\rm phot},i})^2}{2\sigma_{z_{{\rm phot},i}}^2}}\right\}\label{zgauss},
\end{eqnarray}
while if the source has a spectroscopic redshift
\begin{equation}
P_i\, (z_i\, |\, \eta_i) = P_D\, (z_i\, |\, z_{{\rm spec},i}) = \delta(z_i - z_{{\rm spec},i}),
\end{equation}
where $\delta(x)$ is delta's Dirac function. With this distribution we are able to statistically treat all the available information in $z$ space and define the {\it number of companions at redshift $z_1$ in system $j$} as
\begin{equation}
\nu_{j}\,(z_1) = {\rm C}_j\, P_1 (z_1\, |\, \eta_1) \int_{z_{\rm m}^{-}}^{z_{\rm m}^{+}} P_2 (z_2\, |\, \eta_2)\, {\rm d}z_2,\label{nuj}
\end{equation}
where $z_1 \in [z^{-}_j,z^{+}_j]$, the integration limits are
\begin{eqnarray}
z_{\rm m}^{-} = z_1(1-\Delta v^{\rm max}/c) - \Delta v^{\rm max}/c,\\
z_{\rm m}^{+} = z_1(1+\Delta v^{\rm max}/c) + \Delta v^{\rm max}/c,
\end{eqnarray}
the subindex 1 [2] refers to the primary [secondary] galaxy in $j$ system, and the constant ${\rm C}_j$ normalizes the function to the total number of galaxies in the interest range
\begin{equation}
N_{\rm c}^j = \int_{z_j^{-}}^{z_j^{+}} P_1 (z_1\, |\, \eta_i)\, {\rm d}z_1  + \int_{z_j^{-}}^{z_j^{+}} P_2 (z_2\, |\, \eta_2)\, {\rm d}z_2.
\end{equation}
Note that $\nu_j = 0$ if $z_1 < z_j^-$ or  $z_1 > z_j^+$. The function $\nu_j$ (Eq.~[\ref{nuj}]) tells us how the number of close companions in the system $j$, $N_{\rm c}^j$, are distributed in redshift space. The integral in Eq.~(\ref{nuj}) spans those redshifts in which the secondary galaxy has $\Delta v \leq \Delta v^{\rm max}$ for a given redshift of the primary galaxy.

With previous definitions, the number of companions per primary galaxy in the interval $z_{r,k} = [z_k,z_{k+1}]$ is
\begin{equation}
N_{{\rm c},k} = \frac{\sum_j \int_{z_k}^{z_{k+1}}{\nu_j(z_1)}\, {\rm d}z_1}{\sum_i \int_{z_k}^{z_{k+1}} P_i\, (z_i\, |\, \eta_i)\, {\rm d}z_i},\label{ncphot}
\end{equation}
where the index $k$ spans the redshift intervals defined over the redshift range under study. If we integrate over the whole redshift space, $z_{r} = [0,\infty]$, Eq.~(\ref{ncphot}) becomes
\begin{equation}
N_{{\rm c}} = \frac{\sum_j N_{\rm c}^j }{N_1},\label{ncphot2}
\end{equation}
where $N_{\rm c}^j$ is analogous to $N_{\rm c}^i$ in Eq.~(\ref{ncspec}).

\begin{figure}[t]
\centering
\includegraphics{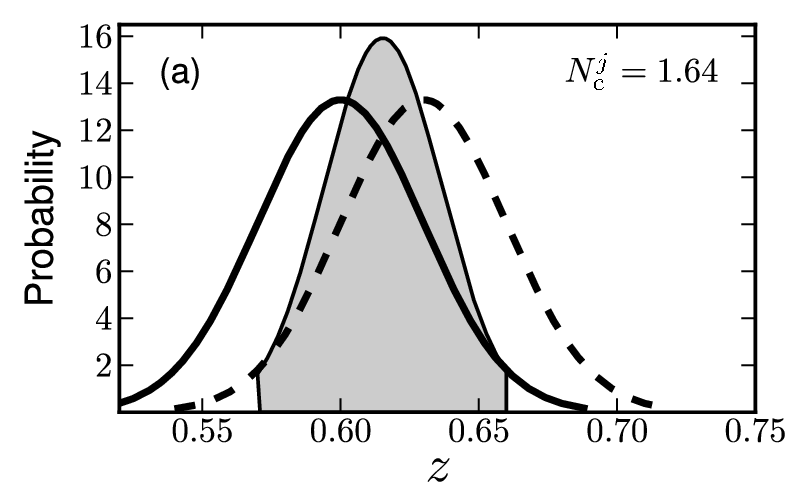}
\includegraphics{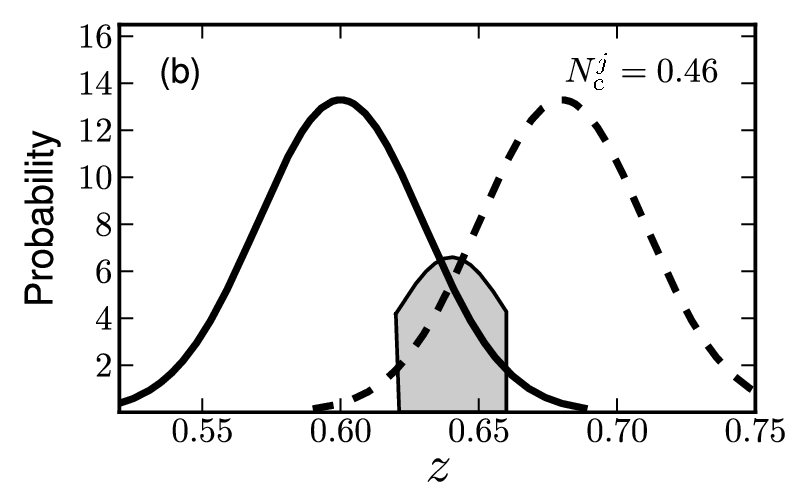}
\includegraphics{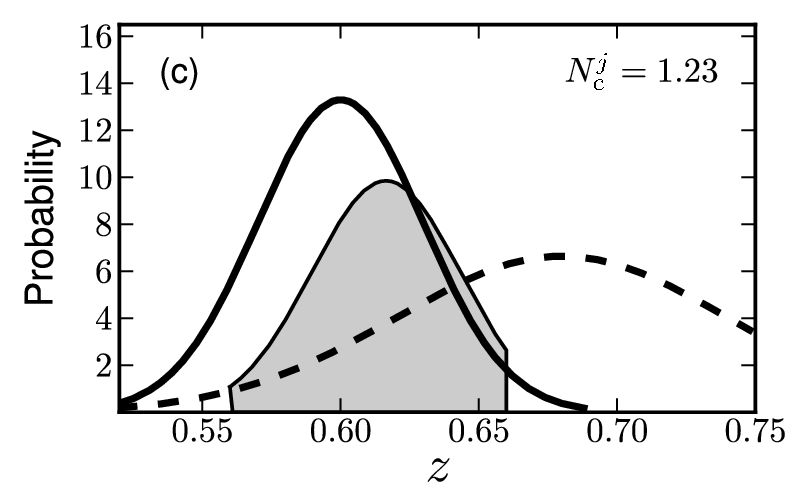}
\caption{Probability distributions in redshift space of a primary galaxy (back solid line), a secondary galaxy (black dashed line), and the function $\nu_j/2$ of the system (grey area). In all panels the primary galaxy has $z_{\rm phot,1} = 0.6$ and $\sigma_{z_{\rm phot,1}} = 0.03$. The secondary galaxy has $z_{\rm phot,2} = 0.63$ and $\sigma_{z_{\rm phot,2}} = 0.03$ in {\it panel (a)}, $z_{\rm phot,2} = 0.68$ and $\sigma_{z_{\rm phot,2}} = 0.03$ in {\it panel (b)}, and $z_{\rm phot,2} = 0.68$ and $\sigma_{z_{\rm phot,2}} = 0.06$ in {\it panel (c)}. The number of close companions in each system, $N_{\rm c}^j$, is labelled in the panels.}
\label{pparfig}
\end{figure}

The definition of function $\nu_j$ in Eq.~(\ref{nuj}) is general, and we can find four different cases in our spectro-photometric samples:

\begin{enumerate}
\item {\it Primary and secondary galaxies have $z_{\rm spec}$.} In this case $\nu_{j}$ is
\begin{equation}
\nu_{j}\,(z_1) = 2 \times \delta (z_1 - z_{\rm spec,1}),
\end{equation}
where $z_{\rm spec,1}$ is the spectroscopic redshift of the more luminous/massive galaxy in the close pair. With this definition, $\nu_{j}$ is not zero only at $z_{\rm spec,1}$. If all galaxies in the sample have $z_{\rm spec}$, Eq.~(\ref{ncphot2}) is equivalent to Eq.~(\ref{ncspec}). These systems provide $N_{\rm c}^j = 2$ companions in Eq.~(\ref{ncphot2}).

\item {\it The primary galaxy has $z_{\rm spec}$ and the secondary $z_{\rm phot}$.} Replacing the corresponding distributions of probability in Eq.~(\ref{nuj}) we obtain
\begin{equation}
\nu_{j}\,(z_1) = {\rm C}_j\, \delta(z_1 - z_{\rm spec,1}) \int_{z_{\rm m}^-}^{z_{\rm m}^+} P_G (z_2\, |\, z_{{\rm phot},2},\sigma_{z_{{\rm phot},2}})\, {\rm d}z_2.
\end{equation}
As in the previous case, $\nu_{j}$ is not zero only at $z_{\rm spec,1}$. These systems provide $N_{\rm c}^j \sim 1$ companions in Eq.~(\ref{ncphot2}).

\item {\it The primary galaxy has $z_{\rm phot}$ and the secondary $z_{\rm spec}$.} In this case Eq.~(\ref{nuj}) becomes
\begin{equation}
\nu_{j}\,(z_1) = {\rm C}_j\, P_G (z_1\, |\, z_{{\rm phot},1},\sigma_{z_{{\rm phot},1}}).
\end{equation}
Function $\nu_{j}(z_1)$ is nonzero in the range
\begin{equation}
\bigg[\frac{z_{{\rm spec},2} - \Delta v^{\rm max}/c}{1+\Delta v^{\rm max}/c},\frac{z_{{\rm spec},2} + \Delta v^{\rm max}/c}{1-\Delta v^{\rm max}/c}\bigg].
\end{equation}
This interval is imposed by the secondary galaxy, and spans those redshifts of the primary galaxy in which the secondary fulfils the condition $\Delta v \leq \Delta v^{\rm max}$. As in the previous case, these systems provide $N_{\rm c}^j \sim 1$ companions in Eq.~(\ref{ncphot2}).

\item {\it Primary and secondary galaxies have $z_{\rm phot}$.} In this case the function $\nu_{j}$ is
\begin{eqnarray}
\nu_{j}\,(z_1) = {\rm C}_j\, P_G (z_1|z_{{\rm phot},1},\sigma_{z_{{\rm phot},1}})\nonumber \\ 
\times \int_{z_{\rm m}^-}^{z_{\rm m}^+} P_G (z_2\, |\, z_{{\rm phot},2},\sigma_{z_{{\rm phot},2}})\, {\rm d}z_2.
\end{eqnarray}
The main difference between this approach and, for example, the one in \citet{ryan08} is that we use the probability distributions of the photometric redshifts to weight the number of companions in each system, $N_{\rm c}^j$, and to minimize projection effects. To illustrate how the weight process works, we show three examples in Fig.~\ref{pparfig}. In all these cases, the primary galaxy has $z_{\rm phot,1} = 0.6$ and $\sigma_{z_{\rm phot,1}} = 0.03$. In the {\it panel (a)}, the secondary galaxy has $z_{\rm phot,2} = 0.63$ and $\sigma_{z_{\rm phot,2}} = 0.03$. With these values the function $\nu_j$ is symmetric and not zero at $[z_j^-,z_j^+] = [0.57,0.66]$. The integral of the function $\nu_j$ over the redshift space gives us the number of companions in the system, which is $N_{\rm c}^j = 1.64$.  In the {\it panel (b)} we increase the redshift of the secondary galaxy to $z_{\rm phot,2} = 0.68$ and keep its previous error. In this example the function $\nu_j$ is nonzero at $[z_j^-,z_j^+] = [0.62,0.66]$, a narrower range than in the previous case. Because of this, the number of companions in this system is only $N_{\rm c}^j = 0.46$. Finally, in the {\it panel (c)} we keep the previous redshift of the secondary and increase its error to $\sigma_{z_{\rm phot,2}} = 0.06$. In this case, the function $\nu_j$ is not symmetric and not zero at $[z_j^-,z_j^+] = [0.56,0.66]$, a similar range to the one in the {\it panel (a)}. However, the secondary galaxy is more extended in redshift space, and the number of companions is lower, $N_{\rm c}^j = 1.23$.
\end{enumerate}

The last step involves normalizing our results to volume-limited samples, by applying the $S_N$ function (Eq.~[\ref{snz}]),
\begin{eqnarray}
\nu_{j}(z_1) = {\rm C}_j\, S_N(z_1) P_1 (z_1\, |\, \eta_1) \int_{z_{\rm m}^-}^{z_{\rm m}^+} S_N(z_2)^{-1} P_2 (z_2\, |\, \eta_2)\, {\rm d}z_2,
\end{eqnarray}
while the final number of companions is
\begin{equation}
N_{{\rm c},k} = \frac{\sum_j \int_{z_k}^{z_{k+1}}{\nu_j(z_1)}\, {\rm d}z_1}{\sum_i \int_{z_k}^{z_{k+1}} S_N(z_i)\, P_i(z_i\, |\, \eta_i)\, {\rm d}z_i},
\end{equation}

In order to estimate the error of $N_{{\rm c},k}$ we use the jackknife technique \citep{efron82}. We compute partial standard deviations, $\delta_j$, for each system $j$ by taking the difference between the measured $N_{{\rm c},k}$ and the same quantity with the $j$th pair removed for the sample, $N_{{\rm c},k}^j$, such that $\delta_j = N_{{\rm c},k} - N_{{\rm c},k}^j$. For a sample with $N$ systems, the variance is given by $\sigma_{N_{{\rm c},k}}^2 = [(N-1) \sum_j \delta_j^2]/N$.

\subsection{Testing the method in a local, volume-limited sample}\label{mgctest}
We tested that our new methodology is able to statistically recover the number of companions in a spectroscopic survey from a photometric one. For this, we study $N_{\rm c}$ in the Millennium Galaxy Catalogue (MGC\footnote{http://eso.org/$\sim$jliske/mgc/}, \citealt{liske03}). This survey comprises 10095 galaxies with $B_{MGC} < 20$ over 37.5~$\deg^2$, with a spectroscopic completeness of 96\% \citep{driver05}. \citet{depropris07} use the MGC to study the number of companions in a volume-limited sample ($0.01 < z < 0.123$, $-21 \leq M_{B} - 5\log h \leq -18$, $N = 3183$ sources), obtaining $N_{\rm c}^{\rm MGC} = 0.035 \pm 0.004$ for $r_{\rm p}^{\rm min} = 0$ and $r_{\rm p}^{\rm max} = 20h^{-1}$ kpc (see also \citealt{depropris05}). We used this volume-limited sample in the present test. In addition, the MGC area had been observed by the Sloan Digital Sky Survey (SDSS\footnote{http://sdss.org/}, \citealt{adelman06}), so every galaxy in the sample also has a photometric redshift. Comparing the $z_{\rm spec}$'s from MGC with the $z_{\rm phot}$'s from SDSS, we obtained $\sigma_{z_{\rm phot}} = 0.02$. We take this uncertainty as representative of SDSS photometric redshifts.

\begin{figure}[t!]
\resizebox{\hsize}{!}{\includegraphics{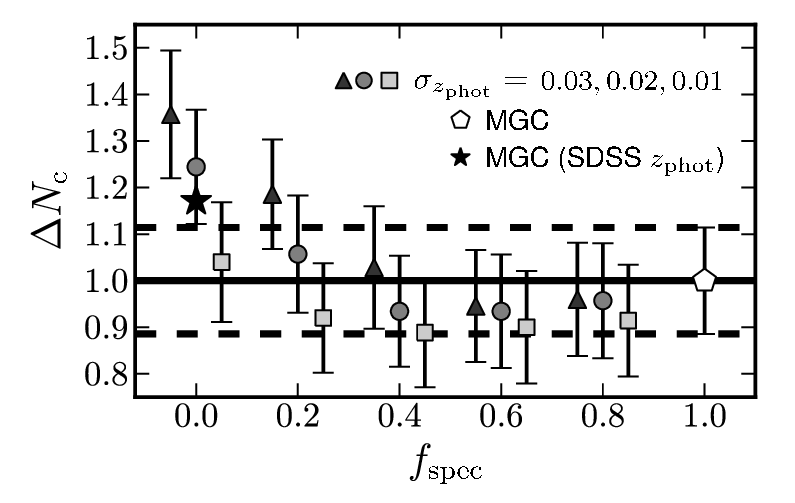}}
\caption{$\Delta N_{\rm c}$ vs $f_{\rm spec}$ for three different values of the photometric redshift uncertainty, $\sigma_{z_{\rm phot}} = 0.03$ (triangles; offset -0.05 for clarity), 0.02 (circles), and 0.01 (squares; offset +0.05 for clarity). We also show the value from MGC spectroscopic sample ($f_{\rm spec} = 1$; white pentagon and black solid line) and its 68\% confidence interval (black dashed lines). The black star is the value obtained with the photometric redshifts of MGC galaxies from SDSS ($f_{\rm spec} = 0$).}
\label{mgcfig}
\end{figure}

We defined $f_{\rm spec}$ as the fraction of the sample's sources with spectroscopic redshift. The MGC sample has $f_{\rm spec} = 1$, while the SDSS sample has $f_{\rm spec} = 0$. To test our method at intermediate $f_{\rm spec}$ and for different $\sigma_{z_{\rm phot}}$, we assigned a $z_{\rm phot}$ to $N(1-f_{\rm spec}$) random sources of the MGC sample, as drawn for a Gaussian distribution (Eq.~[\ref{zgauss}]) with median $z_{\rm spec}$ and a given $\sigma_{z_{\rm phot}}$. Then we measured
\begin{equation}
\Delta N_{\rm c}\,(f_{\rm spec},\sigma_{z_{\rm phot}}) = \frac{N_{\rm c}\,(f_{\rm spec},\sigma_{z_{\rm phot}})}{N_{\rm c}\,(1,0)},
\end{equation}
where $N_{\rm c}\,(f_{\rm spec},\sigma_{z_{\rm phot}})$ is the number of companions in a random sample for a given $f_{\rm spec}$ and $\sigma_{z_{\rm phot}}$, and $N_{\rm c}\,(1,0)$ is the number of companions in the initial spectroscopic sample. We measured $\Delta N_{\rm c}$ for $f_{\rm spec} = 0,0.2,0.4,0.6,0.8$, and $\sigma_{z_{\rm phot}} = 0.01,0.02,0.03$. For each parameter combination, we repeated the process in ten different random catalogues and averaged the values. We summarize the results in Table~\ref{mgctab} and show them in Fig.~\ref{mgcfig}. We find that
\begin{itemize}
\item When we only have spectroscopic information, we obtain the same number of companions as \citet{depropris07}, $N_{\rm c}(1,0) = 0.035 \pm 0.004$. This implies that our methodology is equivalent to those used in spectroscopic samples when $f_{\rm spec} = 1$, as we hoped.

\item When we only have photometric information, our method recovers the initial number of companions (Fig.~\ref{mgcfig}) within error bars where the photometric redshift errors are small ($\sigma_{z_{\rm phot}} = 0.01$). On the other hand, the method overestimates the number of companions for $\sigma_{z_{\rm phot}} = 0.02$ and $0.03$. This implies that we need small photometric redshift errors to avoid projection effects if only photometric information is available.

\item If we use the measured $z_{\rm phot}$'s of MGC sources from SDSS instead of the spectroscopic ones, the number of companions is higher than expected ($N_{\rm c}^{\rm SDSS} = 0.041$) and agrees with that from random samples with $\sigma_{z_{\rm phot}} = 0.02$; that is, our random samples are representative of the observational ones, and we can use them to explore $\Delta N_{\rm c}$ at intermediate values of $f_{\rm spec}$.

\item When our observational errors are not small enough, we need to increase $f_{\rm spec}$ to obtain reliable $N_{\rm c}$ values (Fig.~\ref{mgcfig}). That fixes the redshift of $Nf_{\rm spec}$ galaxies, and we use the photometric information to minimize projection effects. The higher the photometric redshift errors, the higher $f_{\rm spec}$ must be to obtain reliable results. In this test we need $f_{\rm spec} \gtrsim 0.4$ for $\sigma_{z_{\rm phot}} = 0.03$ and $f_{\rm spec} \gtrsim 0.2$ for $\sigma_{z_{\rm phot}} = 0.02$.
\end{itemize}

\begin{table}
\caption{Number of companions in random samples in units of that in the spectroscopic MGC sample, $\Delta N_{\rm c}$}
\label{mgctab}
\begin{center}
\begin{tabular}{lccc}
\hline\hline
$f_{\rm spec}$ & $\sigma_{z_{\rm phot}} = 0.01$ &  $\sigma_{z_{\rm phot}} = 0.02$ &  $\sigma_{z_{\rm phot}} = 0.03$\\
\hline
0      & $1.04 \pm 0.13$ & $1.24 \pm 0.12$ & $1.36 \pm 0.14$ \\
0.2    & $0.92 \pm 0.12$ & $1.06 \pm 0.13$ & $1.19 \pm 0.12$ \\
0.4    & $0.89 \pm 0.12$ & $0.93 \pm 0.12$ & $1.03 \pm 0.13$ \\
0.6    & $0.90 \pm 0.12$ & $0.93 \pm 0.12$ & $0.95 \pm 0.12$ \\
0.8    & $0.91 \pm 0.12$ & $0.96 \pm 0.12$ & $0.96 \pm 0.12$ \\
\hline
\end{tabular}
\end{center}
\end{table}

Summarizing, our new methodology is able to provide reliable $N_{\rm c}$ values from spectro-photometric catalogues for either photometric redshift errors smaller than $\sigma_{z_{\rm phot}}^{\rm max}$ or spectroscopic completeness higher than $f_{\rm spec}^{\rm min}$. For example, in our local test we find $\sigma_{z_{\rm phot}}^{\rm max} = 0.01$, while $f_{\rm spec}^{\rm min}$ depends on the photometric redshift errors. These limits are not applicable to the GOODS-S catalogue because of the different galaxy number densities (e.g., the MGC volume-limited sample contains three clusters, \citealt{driver05}; while we remove the most prominent LSS in GOODS-S, see Sect.~\ref{data}) or the different search areas in the sky plane, which decreases by a factor of 15 from $z = 0.1$ to $z = 0.7$ for the same physical area. Photometric redshifts errors in our GOODS-S catalogue are typically $\sigma_{z_{\rm phot}} > 0.05$, while $f_{\rm spec} \gtrsim 0.4$ (see Sect.~\ref{fspec}). Our first step is therefore to compare our results at $0.2 \leq z < 1.1$ with those from spectroscopic samples to ensure that we can obtain reliable $N_{\rm c}$ values from the GOODS-S catalogue. We perform this comparison in Sects.~\ref{ncrp} and \ref{ncb}.

\subsection{Number of companions as a function of $r_{\rm p}^{\rm max}$}\label{ncrp}
In this section we use the methodology developed in Sect.~\ref{ncp} to studying the variation in $N_{\rm c}$ with $r_{\rm p}^{\rm max}$. This is a consistency test of our method because $N_{\rm c} \propto r_{\rm p}^{3 - \gamma}$ \citep{patton02,lin08,deravel09}, where $\gamma$ is the exponent in the correlation function, $\xi(r) \propto (r_0/r)^{\gamma}$. The $\gamma$ values in the literature are consistent with $\gamma \sim 1.7$ \citep[e.g.,][]{lefevre05}, so we expect $N_{c} \propto r_{\rm p}^{1.3}$.

In this test we define two samples: one comprises all the galaxies in our catalogue with $-22 < M_B \leq -19$ and $0.2 < z \leq 1.1$, named $G_{\rm phot}$, while another comprises only the galaxies of $G_{\rm phot}$ with $z_{\rm spec}$, named $G_{\rm spec}$. We take these two samples as primary and secondary, and we do not impose any difference in luminosity between both galaxies when searching for close companions. We vary $r_{\rm p}^{\rm max}$ from $20h^{-1}$ kpc to $70h^{-1}$ kpc in $5h^{-1}$ kpc steps, and include the radius $21h^{-1}$ kpc. In all cases we use $r_{\rm p}^{\rm min} =  6h^{-1}$ kpc.

\begin{figure}[t!]
\resizebox{\hsize}{!}{\includegraphics{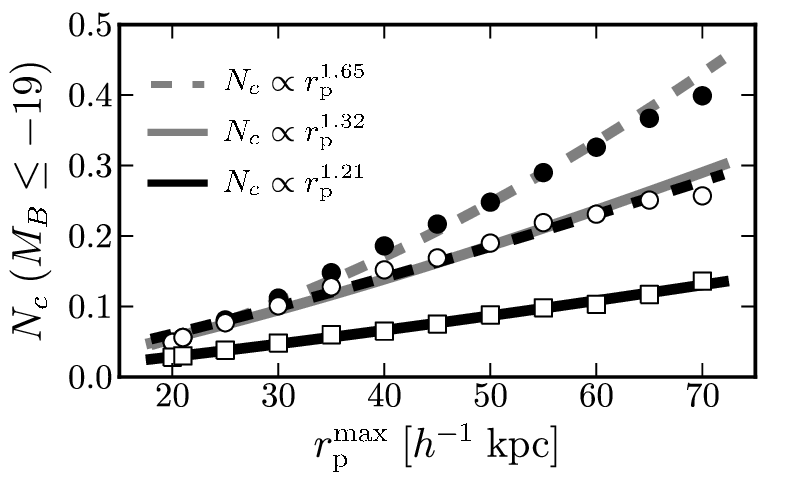}}
\caption{Number of companions vs $r_{\rm p}^{\rm max}$ for galaxies with $-22 < M_B \leq -19$ and $0.2 < z \leq 1.1$. White dots are obtained using approach 1 in the multiple-system treatment ($G_{\rm phot,1}$), while black dots are from approach 2 ($G_{\rm phot,2}$, see text for details). White squares are from approach 1 over sources with spectroscopic redshift ($G_{\rm spec,1}$). The black solid line is the best power-law fit, $N_\mathrm{c} \propto r_{\rm p}^{q}$, to the $G_{\rm spec,1}$ data, grey solid line to the $G_{\rm phot,1}$ data, and grey dashed line to the $G_{\rm phot,2}$ data. The exponent of the fits is labelled in the figure. The black dashed line is the expected number of companions in the  $G_{\rm phot}$ sample, derived from $G_{\rm spec,1}$ data.}
\label{ncrpfig}
\end{figure}

When we increase the radius of search, we start to find two or more secondary galaxies close to each primary galaxy. We treat these multiple systems with two different approaches:

\begin{enumerate}
\item We study all the possible pairs as independent systems; that is, in a multiple system that comprises the galaxies A, B, and C; we study the subsystems A-B, A-C, and B-C.
\item We only study {the most massive pair}, that is, the one with the lowest difference in luminosity/mass between the primary and the secondary galaxies. If this spatial pair is not a close pair in redshift space, we study the next more representative pair, and so on.
\end{enumerate}
We applied approach 1 to the $G_{\rm spec}$ sample, while applying both to the $G_{\rm phot}$ sample, called $G_{\rm phot,1}$ and $G_{\rm phot,2}$, respectively. We did this because approach 1 can fail in the $G_{\rm phot}$ sample owing to the high uncertainties in $z_{\rm phot}$, making a projection correction necessary, but this is not the case if all the sources in the sample have $z_{\rm spec}$. Approach 2 is more conservative and we can miss, for example, a minor companion of a galaxy with a major one, but mitigate the contamination by projection effects. We summarize the results in both Table~\ref{ncrptab} and Fig.~\ref{ncrpfig}. Fitting a power-law function to the data, $N_\mathrm{c} \propto r_{\rm p}^q$, we obtained $q = 1.65$ in the $G_{\rm phot,1}$ case, $q = 1.32$ in the $G_{\rm phot,2}$ case, and $q = 1.21$ in the $G_{\rm spec,1}$ case. We see that

\begin{itemize}
\item The trends obtained in the $G_{\rm spec,1}$ and $G_{\rm phot,2}$ cases agree with the expected $q \sim 1.3$; however, the value of $q$ in the $G_{\rm phot,1}$ case is higher than expected, reflecting the increased contamination by projection effects when the search area increases.

\item For a given $r_{\rm p}^{\rm max}$, one expects that $N_\mathrm{c} \propto n_{g}$, where $n_{g}$ is the number of galaxies in the sample \citep{lin08}; in our case, $n_{g}(G_{\rm spec}) = 0.47 n_{g}(G_{\rm phot})$, so we can predict the expected number of companions in the $G_{\rm phot}$ sample from the obtained in the $G_{\rm spec,1}$ (Fig.~\ref{ncrpfig}). We see that the expected number is in excellent agreement with the $G_{\rm phot,2}$ number of companions.
This implies that our methodology is able to recover statistically, from the GOODS-S spectro-photometric sample, the same $N_\mathrm{c}$ that in a fully spectroscopic one, as we hoped (see also Sects.~\ref{mgctest} and \ref{fspec}).

\item The number of companions at $r_{\rm p}^{\rm max} < 25h^{-1}$ kpc in the $G_{\rm phot,1}$ and $G_{\rm phot,2}$ cases are the same. This implies that projection effects start to be important at $r_{\rm p}^{\rm max} \gtrsim 30h^{-1}$ kpc.
\end{itemize}

In this section we have applied our new methodology to the GOODS-S catalogue with positive results, and in the following we use approach $2$ to treat multiple systems. Note that 47\% of the $G_{\rm phot}$ sources have $z_{\rm spec}$. We study how the spectroscopic completeness of the sample affect our results in Sect.~\ref{fspec} (see also Sect~\ref{mgctest}).

\begin{table}
\caption{Number of companions as a function of $r_{\rm p}^{\rm max}$ for approaches 1 (independent pairs) and 2 (more representative pair) in the multiple-system treatment.}
\label{ncrptab}
\begin{center}
\begin{tabular}{lccc}
\hline\hline
$r_{\rm p}^{\rm max}$ & \multicolumn{3}{c}{$N_{\rm c} (0.2 \leq z < 1.1, -22 < M_B \leq -19)$}\\
\noalign{\smallskip}
\cline{2-4}
\noalign{\smallskip}                      
  ($h^{-1}$ kpc)     & $G_{{\rm phot},1}$ &  $G_{{\rm phot},2}$ &  $G_{{\rm spec},1}$\\
\hline
20    & $0.049 \pm 0.006$ & $0.049 \pm 0.006$ & $0.028 \pm 0.008$ \\
21    & $0.056 \pm 0.006$ & $0.056 \pm 0.006$ & $0.030 \pm 0.008$ \\
25    & $0.081 \pm 0.007$ & $0.077 \pm 0.007$ & $0.038 \pm 0.009$ \\
30    & $0.112 \pm 0.008$ & $0.101 \pm 0.008$ & $0.048 \pm 0.010$ \\
35    & $0.148 \pm 0.009$ & $0.128 \pm 0.008$ & $0.060 \pm 0.011$ \\
40    & $0.186 \pm 0.009$ & $0.152 \pm 0.009$ & $0.065 \pm 0.012$ \\
45    & $0.217 \pm 0.009$ & $0.169 \pm 0.009$ & $0.075 \pm 0.013$ \\
50    & $0.248 \pm 0.009$ & $0.190 \pm 0.009$ & $0.088 \pm 0.013$ \\
55    & $0.290 \pm 0.009$ & $0.219 \pm 0.009$ & $0.098 \pm 0.014$ \\
60    & $0.326 \pm 0.008$ & $0.231 \pm 0.009$ & $0.103 \pm 0.014$ \\
65    & $0.367 \pm 0.008$ & $0.251 \pm 0.009$ & $0.117 \pm 0.015$ \\
70    & $0.399 \pm 0.008$ & $0.257 \pm 0.009$ & $0.136 \pm 0.016$ \\
\hline
\end{tabular}
\end{center}
\end{table}

\subsection{Border effects in redshift and in the sky plane}\label{border}
When we search for a primary source companion, we define a volume in the sky plane-redshift space. If the primary source is near the boundaries of the survey, a fraction of the search volume lies outside of the effective volume of the survey. To account for this, we use a correction factor $f_b$, the fraction of the area $\pi [(r_{\rm p}^{\rm max})^2 - (r_{\rm p}^{\rm min})^2]$ around the primary galaxy that lies in the survey area. This factor is positive and depends on $z$ because the projected distance in the sky is function of primary source redshift. We studied how this factor affects the previous section results, finding that border effects are representative (i.e., $1\sigma$ discrepancy) only at $r_{\rm p}^{\rm max} \gtrsim 70h^{-1}$ kpc.

We avoid the redshift incompleteness by including in the samples not only the sources inside the redshift range under study, but also those sources with $z_i + 2\sigma_i > 0.2$ and $z_i - 2\sigma_i < 1.1$ (see Sects.~\ref{ncrp} and \ref{ncb}). This implies that our samples comprise sources with $z \in [0.1,1.3)$.

\begin{figure*}[t]
\centering
\includegraphics[width = 8.5cm]{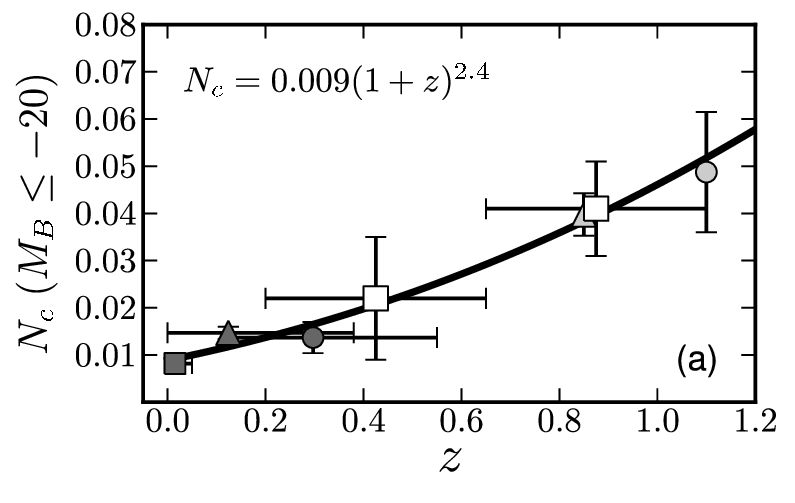}
\includegraphics[width = 8.5cm]{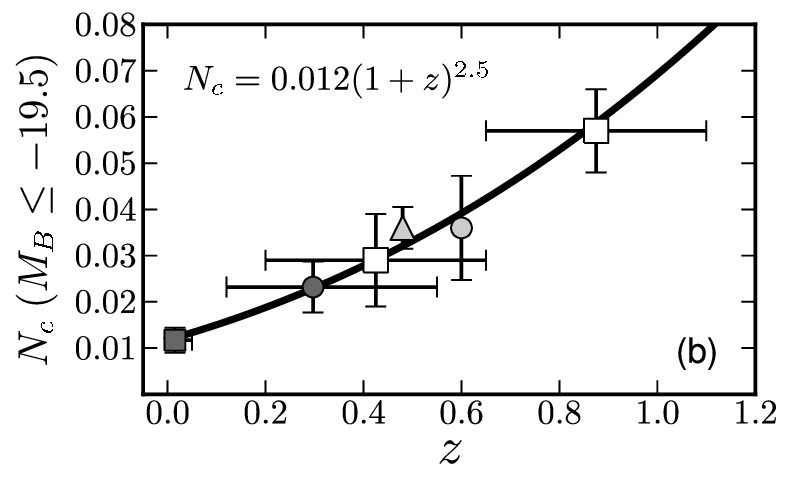}
\includegraphics[width = 8.5cm]{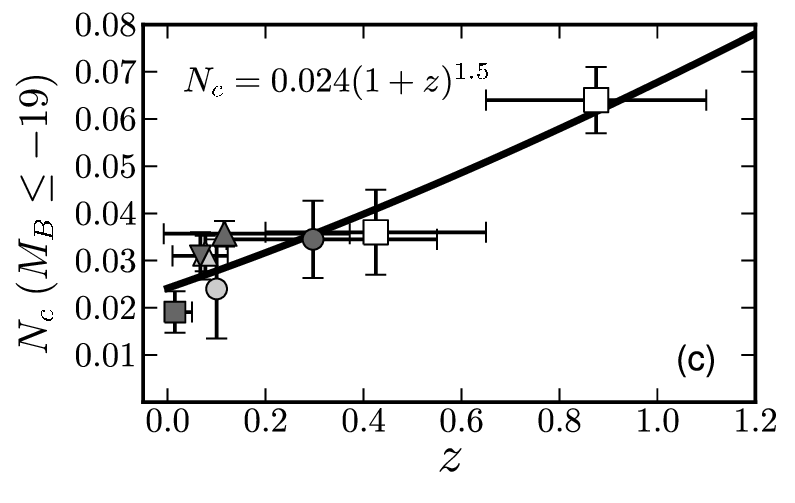}
\includegraphics[width = 8.5cm]{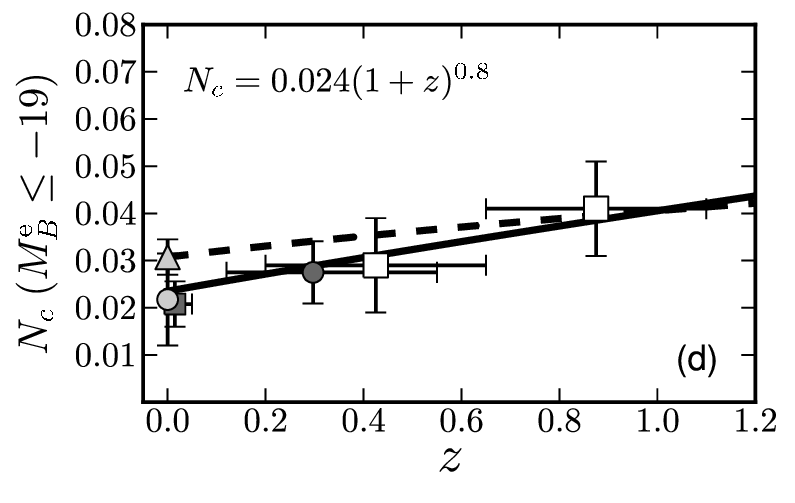}
\caption{Number of companions vs redshift for different $B$-band luminosity selections: {\it panel (a)} for $-22 < M_B \leq -20$ galaxies, {\it panel (b)} for $-22 < M_B \leq -19.5$ galaxies, {\it panel (c)} for $-22 < M_B \leq -19$ galaxies, and {\it panel (d)} for $-22 < M_B^{\rm e} (AB) \leq -19$ galaxies (see text for details). Symbols are the number of companions from this work (GOODS-S, white squares; MGC, white triangles), \citet[dark grey squares]{patton00}, \citet[dark grey circles]{patton02}, \citet[light grey circles]{lin04}, and \citet[light grey triangles]{lin08}. Errors bars in $z$ axis show the redshift range spans for each point. Black solid lines in all the panels are the best least-squares fit of a power-law function to the data. The parameters of the fit are in each panel. Black dashed line in {\it panel (d)} is the power-law parametrization provide by \citet{lin08} with $m = 0.4$. To facilitate comparison, the scales are the same in all the panels.}
\label{ncbfig}
\end{figure*}

\begin{table*}
\caption{Number of companions in $B$-band luminosity-selected samples}
\label{ncbtab}
\begin{center}
\begin{tabular}{lcccc}
\hline\hline
Reference & $z$ & $-22 < M_B \leq -20$ & $-22 < M_B \leq -19.5$ & $-22 < M_B \leq -19$\\
\hline
Present work (GOODS-S)  & 0.425  & $0.022 \pm 0.013$ & $0.029 \pm 0.010$ & $0.036 \pm 0.009$ \\
			& 0.875  & $0.041 \pm 0.010$ & $0.057 \pm 0.009$ & $0.064 \pm 0.008$\\
Present work (MGC)	& 0.092  & $0.011 \pm 0.004$ & $0.024 \pm 0.005$ & $0.034 \pm 0.005$ \\
\citet{patton00} (a)	& 0.015  & $0.008 \pm 0.002$ & $0.012 \pm 0.003$ & $0.019 \pm 0.004$ \\
\citet{patton02} (b)    & 0.297  & $0.014 \pm 0.003$ & $0.023 \pm 0.006$ & $0.035 \pm 0.008$ \\
\citet{lin04} (c)     	& 0.1    & \ldots            & \ldots            & $0.024 \pm 0.010$ \\
		    	& 0.6    & \ldots            & $0.036 \pm 0.011$ & \ldots            \\
		    	& 1.1    & $0.049 \pm 0.013$ & \ldots            & \ldots            \\
\citet{lin08} (c)       & 0.077  & \ldots            & \ldots            & $0.031 \pm 0.004$ \\
		    	& 0.48   & \ldots            & $0.036 \pm 0.005$ & \ldots            \\
	            	& 0.85   & $0.040 \pm 0.005$ & \ldots            & \ldots           \\
\hline
\end{tabular}
\end{center}
\begin{footnotetext}
TNOTES. (a) Original data are obtained with $h = 1$. (b) Original data are obtained with $h = 1$ and $M_B^{\rm e}(z) = M_B - z$ sample selection. (c) We use the Eq.~(\ref{ncz}) parameters provide for $-21 < M_B^{\rm e}(z)$ (AB) $\leq -19$ galaxies, where $M_B^{\rm e}(z) = M_B - Qz$, with $Q = 1$ in \citet{lin04} and $Q = 1.3$ in \citet{lin08}. The search radius is $10h^{-1} < r_{\rm p} \leq 30h^{-1}$ kpc, so we multiply their values by $3/4$ to normalize their results at $15h^{-1}$ kpc (from the correlation funtion study of \citealt{bell06}).
\end{footnotetext}
\end{table*}

\section{Results}
\subsection{Number of close companions in $B$-band luminosity-selected samples}\label{ncb}
In this section we study the number of companions as a function of $z$ and $M_B$. For this, we define three luminosity-selected samples, $-22 < M_B\leq -20$, $-22 < M_B\leq 19.5$, and $-22 < M_B\leq 19$, and study $N_\mathrm{c}$ in two redshift ranges, named $z_{r,1} = [0.2,0.65)$ and $z_{r,2} = [0.65,1.1)$. We chose these redshift ranges to ensure good statistics in our study. In addition, we take the three previous luminosity-selected samples as primary and secondary, and do not impose any limit in primary to secondary luminosity ratio. We apply the same luminosity selections and search parameters to the MGC sample described in Sect.~\ref{mgctest} to obtain consistent data points at $z = 0.092$. We summarize our results with those from previous spectroscopic studies in literature in Table~\ref{ncbtab} and show them in Fig.~\ref{ncbfig}. We find that our results from spectro-photometric samples are in excellent agreement with those from fully spectroscopic ones when similar search parameters (i.e., luminosity selection and radius range) are applied.

The dependence of $N_\mathrm{c}$ on $z$ can be parametrized with a power-law function,
\begin{equation}
N_{\rm c}(z) = N_{\rm c}(0)(1+z)^{m}.\label{ncz}
\end{equation}
The least-squares fits of Eq.~(\ref{ncz}) to the data are summarized in Table~\ref{ncbfit}. We find that the number of companions evolves faster for more luminous galaxies. The index $m$ decreases from  $m = 2.5\pm0.1$ for $M_B < -20$ galaxies to $m = 1.6\pm0.4$ for $M_B < -19$ galaxies. This trend is well described as $m \propto (-0.8\pm0.4) \times M_{B}^{\rm sel}$. On the other hand, $N_{\rm c}(0)$ increases when luminosity decreases, in agreement with \citet{patton00,patton02}; and \citet{lin04} results. These trends also agree with those predicted by the cosmological model of \citet{khochfar01}.

In the near universe ($z \sim 0.1$), \citet{patton08} study the dependence of $N_{\rm c}$ on $r$-band luminosity in an SDSS sample. They find a nearly constant value over the range $-22 < M_{r} < -18$ for $r_{\rm p}^{\rm min} = 5h^{-1}$ kpc and $r_{\rm p}^{\rm max} = 20h^{-1}$ kpc close companions, $N_{\rm c} = 0.021\pm0.001$, in contrast to our measured evolution with luminosity of $N_{\rm c}(0)$. However, they only account for major companions ($\Delta M_r = 0.75$), while we do not impose any luminosity constraint. This implies that we are sensitive to minor companions, more numerous than major (see Sect.~\ref{ncmass}), in the lower luminosity sample, that lead to an increase in $N_{\rm c}$. On the other hand, \cite{deravel09} measure the evolution of the major merger fraction ($\Delta M_B = 1.5$) in the VIMOS-VLT Deep Survey (VVDS\footnote{http://www.oamp.fr/virmos/vvds.htm}, \citealt{lefevre05}) up to $z \sim 1$, finding that the index $m$ is lower in more luminous samples, an opposite trend to what we find. This discrepancy can be explained again by the different definition of companion: we lose major companions near the selection luminosity, but gain the minor ones of the more luminous galaxies. These two examples point out that similar definitions of companion are needed to compare results from different studies and surveys.

\citet{lin08} study the evolution of $N_\mathrm{c}$ at $0 < z < 1.2$ from $\sim35000$ spectroscopic sources, finding $N_{\rm c}(0) = 0.031 \pm 0.004$ and $m = 0.4\pm0.2$, a low value that is incompatible with those in the present work. However, \citet{lin08} galaxies have $-21 < M_B^{\rm e}(z) {\rm (AB)} \leq -19$, where $M_B^{\rm e}(z) = M_B - 1.3z$, a different selection than ours. To explore whether this can be the origin of the discrepancy in $m$ value, as suggested by \citet{kar07}, we mimic the \citet{lin08} selection in the literature and in present work as close as possible. In the latter, the \citet{lin08} selection at $z = 0.425$ is $M_B \lesssim -19.5$ and is $M_B \lesssim -20$ at $z = 0.875$. We summarize the data in Table~\ref{ncblin} and show them in the panel (d) of Fig.~\ref{ncbfig}. We obtain $m = 0.7 \pm 0.4$, a low value compatible with the \citet{lin08} result: the selection of the sample is a key issue in determining $m$, and must be taken into account when one compares different works. Finally, the models of \citet{berrier06} predict $m = 0.4-1.0$, in good agreement with our result.

\subsubsection{Dependence on the spectroscopic completeness of the sample}\label{fspec}
In previous section we show that our methodology provides reliable $N_{\rm c}$ values, compatible with those obtained in fully spectroscopic samples, from GOODS-S spectro-photometric samples. As we show in Sect.~\ref{mgctest}, this can stem from either i) small photometric redshift errors or ii) enough spectroscopic completeness. The $-22 < M_B\leq -20$ sample has $f_{\rm spec} = 0.54$, where $f_{\rm spec}$ is the fraction of sources with $z_{\rm spec}$ in the sample, the $-22 < M_B\leq -19.5$ sample has $f_{\rm spec} = 0.49$, and the $-22 < M_B\leq -19$ sample has $f_{\rm spec} = 0.42$. This means that our methodology works at least for $f_{\rm spec} \gtrsim 0.4$ samples. To check that photometric redshift errors in GOODS-S catalogue are small enough to skip spectroscopic information, we repeat the study in the previous section but using $z_{\rm phot}$ for all the sources, although some have $z_{\rm spec}$ (i.e., $f_{\rm spec} = 0$). Unfortunately, we obtain {\it higher values of $N_{\rm c}$ than expected}. This implies (i) that we need $f_{\rm spec} \gtrsim 0.4$ in the current GOODS-S samples to avoid an overestimation in $N_{\rm c}$. because of projection effects and (ii) that lower $\sigma_{\delta_{z}}$ are needed in order to apply our methodology to $f_{\rm spec} = 0$ catalogues up to $z \sim 1$ (e.g., the COSMOS\footnote{Cosmological Evolution Survey, \citealt{scoville07} (http://cosmos.astro.caltech.edu/index.html).} survey, where $\sigma_{\delta_{z}} \sim 0.01$, \citealt{ilbert09cos}).

\begin{table}
\caption{Fit parameters of the function $N_{\rm c}(z) = N_{\rm c}(0)(1+z)^m$ to the data}
\label{ncbfit}
\begin{center}
\begin{tabular}{lcc}
\hline\hline
Sample selection & $N_{\rm c}(0)$  & $m$\\
\hline
$-22 < M_B \leq -20$     &  $0.008 \pm 0.001$ & $2.5 \pm 0.1$ \\
$-22 < M_B \leq -19.5$   &  $0.014 \pm 0.002$ & $2.2 \pm 0.3$ \\
$-22 < M_B \leq -19$     &  $0.023 \pm 0.003$ & $1.6 \pm 0.4$ \\
\hline
\end{tabular}
\end{center}
\end{table}

\begin{table}
\caption{Number of companions for $-21 < M_B^{\rm e} {\rm (AB)} \lesssim -19$ galaxies}
\label{ncblin}
\begin{center}
\begin{tabular}{lcc}
\hline\hline
Reference & $z$ & $N_\mathrm{c}$\\
\hline
\citet{lin04}       	& 0            	&$0.022 \pm 0.010$\\
\citet{lin08}       	& 0            	&$0.031 \pm 0.004$\\
\citet{patton00}	& 0.015        	&$0.021 \pm 0.005$\\
Present work (MGC)	& 0.092		&$0.034 \pm 0.005$\\
\citet{patton02}	& 0.297        	&$0.028 \pm 0.007$\\
Present work (GOODS-S) 	& 0.425 	&$0.029 \pm 0.013$\\
Present work (GOODS-S)	& 0.875  	&$0.041 \pm 0.010$\\
\hline
\end{tabular}
\end{center}
\end{table}

\subsection{Number of close companions in mass-selected samples}\label{ncmass}
In this section we study the number of major and minor companions of primary galaxies with $\log(M_{\star,1}^{\rm sel}/M_{\odot}) = 10$. As in previous sections, we define two redshift ranges, named $z_{r,1} = [0.2,0.65)$ and $z_{r,2} = [0.65,1.1)$. In the following we denote the galaxy mass ratio as
\begin{equation}
\mu \equiv \frac{M_{\star,2}}{M_{\star,1}},
\end{equation}
where $M_{\star,1}$ and $M_{\star,2}$ are the stellar mass of the primary and the secondary galaxies in the pair, respectively. To explore the dependence of $N_{\rm c}$ on $\mu$, we vary the mass ratio under study from $\mu \geq 1/2$ to $\mu \geq 1/10$. We take major companions, denoted $N_{\rm c}^{\rm M}$, as those with $\mu \geq 1/3$, while minor companions, denoted $N_{\rm c}^{\rm m}$, have $\mu \geq 1/10$. With these definitions, major companions are included in $N_{\rm c}^{\rm m}$. For completeness, we take $M_{\star,2}^{\rm sel} = \mu \times M_{\star,1}^{\rm sel}$ hereafter, unless noted otherwise. All the mass-selected samples in this section have $f_{\rm spec} \geq 0.4$ (see Sect.\ref{fspec} for details), varying from $f_{\rm spec} = 0.65$ for $M_{\star} \geq 10^{10}\ M_{\odot}$ galaxies to $f_{\rm spec} = 0.41$ for $M_{\star} \geq 10^{9}\ M_{\odot}$ galaxies.

\begin{figure}[t!]
\resizebox{\hsize}{!}{\includegraphics{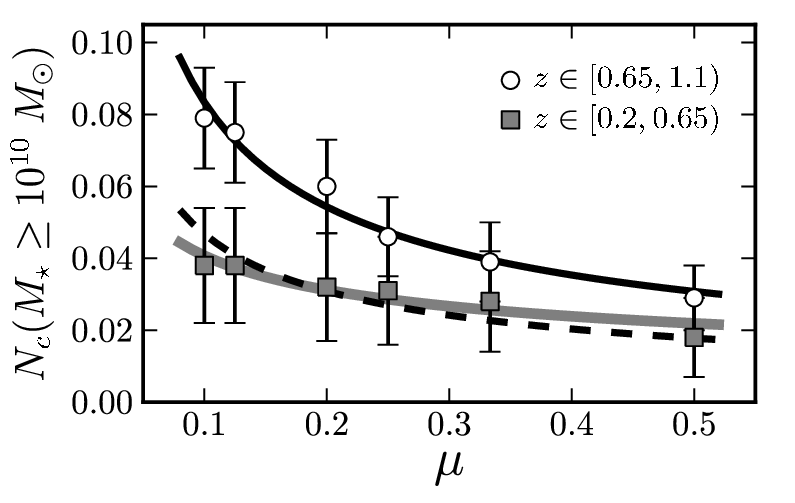}}
\caption{Number of companions vs mass ratio $\mu$ for primary galaxies with $M_{\star} \geq 10^{10}\ M_{\odot}$. Open circles are for $z = 0.875$ galaxies, and grey squares for $z = 0.425$ galaxies. The solid lines are the best fit of a power-law function, $N_\mathrm{c} \propto \mu^{s}$, to the data. The black dashed line is the best fit of a power-law function with a fixed exponent $s = -0.6$ to the $z = 0.425$ data (see text for details).}
\label{ncRfig}
\end{figure}

We summarize our results in Table~\ref{ncmasssectab} and show them in Fig.~\ref{ncRfig}. We find that (i) the number of close companions decreases with redshift for every $\mu$. Combining all the $\mu$ values, this evolution is described by $N_{\rm c}(z) \propto (1+z)^{2.1\pm0.5}$ well. This evolution agrees with the finding by \citet{deravel09} for the major merger ($\mu \geq 1/4$) fraction of $M_{\star} \geq 10^{10}\ M_{\odot}$ galaxies in VVDS-Deep survey, $m = 2.04$. And (ii), the number of close companions grows when $\mu$ decreases. Combining both redshift ranges, we obtain $N_{\rm c}^{\rm m} = (1.7\pm0.3) \times N_{\rm c}^{\rm M}$; that is, {\it the number of minor companions is roughly twice the number of major companions}. 

Only a few works have studied minor mergers statistically. \citet{jogee09} report the minor merger ($\mu \gtrsim 1 / 10$) fraction in Galaxy Evolution from 
Morphology and SEDs (GEMS\footnote{http://www.mpia.de/GEMS/gems.htm}, \citealt{rix04}) for $M_{\star} \gtrsim 4 \times 10^{10}\ M_{\odot}$ galaxies (\citealt{salpeter55} IMF), and estimate that minor mergers are three times major ($\mu \gtrsim 1 / 4$) mergers. \citet{lotz08ff} use $G-M_{20}$ morphological indices to determine the minor merger ($\mu \gtrsim 1 / 9$, \citealt{lotz09t}) fraction in All-Wavelength Extended Groth Strip International Survey (AEGIS\footnote{http://aegis.ucolick.org/}, \citealt{davis07}). Their values are in good agreement with the \citet{jogee09} results, but they do not estimate the major merger fraction. In the local Universe, \citet{darg10i} estimate that minor mergers are twice major mergers in Galaxy Zoo\footnote{http://www.galaxyzoo.org} \citep{lintott08}; the latter is based on the visual classification of SDSS galaxies by internet users. Finally, \citet{woods07} study the different properties of major ($\Delta m_{z} < 2$) and minor ($\Delta m_{z} > 2$) close pairs in SDSS, which corresponds to $\mu \sim 1/7$. Unfortunately, they do not attempt to derive merger fractions, but the influence of close companions on galaxy properties (see also \citealt{ellison08}). Summarizing, literature values are consistent with $N_{\rm c}^{\rm m} = 2-3 \times N_{\rm c}^{\rm M}$. The different methodologies (close pair vs morphology) and sample selections make quantitative comparisons difficult (see also Sect.~\ref{mmrm}), but the qualitative agreement is remarkable.

The observed dependence of $N_{\rm c}$ on $\mu$ is parametrized well as $N_{\rm c}(\mu) \propto \mu^{s}$, as predicted by the simulations of \citet{maller06}. Fitting a power-law function to the data we obtain $s = -0.6 \pm 0.2$ at $z = 0.875$, and $s = -0.4 \pm 0.2$ at $z = 0.475$. However, the observed evolution in $s$ is not significant: if we impose $s = -0.6$ in the range $z_{r,1}$, the observational values are also described well within the uncertainties. The highest difference between the model with fixed $s = -0.6$ and the observations occurs at $\mu \geq 1/10$, i.e., for minor companions. This suggests that  $\log(M_{\star}/M_{\odot}) \geq 10$ galaxies have accreted satellite galaxies between $z \sim 0.9$ and $z \sim 0.5$. Since most of these low-mass satellites are expected to be gas-rich, this accretion could explain the residual star formation observed in early-types galaxies at $z \sim 0.6$ \citep{kaviraj10}. Studies with larger samples are needed to constraint the differential evolution with redshift, if any, of the number of major and minor companions.

\begin{table}
\caption{Number of close companions of $M_{\star,1} \geq 10^{10}\ M_{\odot}$ galaxies as a function of mass ratio $\mu$}
\label{ncmasssectab}
\begin{center}
\begin{tabular}{lccc}
\hline\hline
Secondary sample & $\mu$ & $z = 0.425$ & $z = 0.875$\\
\hline
$\log (M_{\star,2}/M_{\odot}) \geq 9.7$ & 1/2  & $0.018 \pm 0.011$ & $0.029 \pm 0.009$ \\
$\log (M_{\star,2}/M_{\odot}) \geq 9.5$ & 1/3  & $0.028 \pm 0.014$ & $0.039 \pm 0.011$ \\
$\log (M_{\star,2}/M_{\odot}) \geq 9.4$ & 1/4  & $0.031 \pm 0.015$ & $0.046 \pm 0.011$ \\
$\log (M_{\star,2}/M_{\odot}) \geq 9.3$ & 1/5  & $0.032 \pm 0.015$ & $0.060 \pm 0.013$ \\
$\log (M_{\star,2}/M_{\odot}) \geq 9.1$ & 1/8  & $0.038 \pm 0.016$ & $0.075 \pm 0.014$ \\
$\log (M_{\star,2}/M_{\odot}) \geq 9$   & 1/10 & $0.038 \pm 0.016$ & $0.079 \pm 0.014$ \\
\hline
\end{tabular}
\end{center}
\end{table}

\section{Major merger rate: close pairs vs morphological criteria}\label{mrpair}
The local study ($z \sim 0.09$) of \cite{depropris07} shows that merger fractions by close pairs and by morphological criteria (i.e., taking highly distorted galaxies as major merger remnants) give similar merger rates when samples are compared carefully. However, in the range $0.2 < z < 1.2$ both methods yield different merger rates, some times by an order of magnitude \citep[e.g.,][]{lin04}. In this section we compare the major merger rate inferred by our close pair study, $\Re_{\rm M}^{\rm pair}$, with that from \citet[][L09 hereafter]{clsj09ffgoods} by morphological criteria, $\Re_{\rm M}^{\rm mph}$. In L09 gas-rich major merger remnants are selected as those galaxies with high values of the asymmetry index ($A$, \citealt{abraham96,conselice03}). L09 determine the morphological merger fraction of $\log(M_{\star}/M_{\odot}) \geq 10$ galaxies in the same GOODS-S catalogue that we use in the present paper, and lead with three important sources of systematics: (i) they avoid morphological $K$-corrections by measuring the asymmetries in the rest-frame $B$-band; (ii) the deal with the loss of information with redshift (i.e., spatial resolution descent and cosmological dimming) by artificially redshifting all the galaxies to a unique and representative redshift, $z_{\rm d} = 1$; and (iii) they take the effect of observational errors in $z$ and $A$ into account by maximum likelihood techniques developed in \citet{clsj08ml}. To obtain $\Re_{\rm M}^{\rm mph}$ from the merger fraction, they assume \citet{pgon08} mass functions, as in the present paper, and a typical timescale of $T_A \sim 0.5$ Gyr \citep{conselice06ff,conselice09t,lotz08t,lotz09t}.

Following \citet{lin04}, we define the major merger rate by close pairs as
\begin{equation}
\Re_{\rm M}^{\rm pair}(z,M_{\star}) = \frac{1}{2} \frac{4}{3} C_{\rm m}\, \rho(z,M_{\star})\, N_{\rm c}^{\rm M}(z,M_{\star})\, T_{\rm pair}^{-1},\label{mrpar}
\end{equation}
where the factor 1/2 is for obtaining the number of merger systems from the number of close companions, the factor 4/3 takes the lost companions in the inner $6h^{-1}$ kpc into account \citep{bell06}, $C_{\rm m}$ accounts for the fraction of the observed systems that really merge in $T_{\rm pair}$, and $\rho(z,M_{\star})$ is the comoving number density of galaxies more massive than $M_{\star}$ at redshift $z$. We take $C_{\rm m} = 0.6\pm0.1$ \citep{patton00,bell06,lin08,deravel09}, $T_{\rm pair} = 0.75 \pm 0.25$ Gyr \citep{kit08,deravel09}, and determine $\rho(z,M_{\star})$ with the mass functions from \citet{pgon08}.

Although L09 determine $\Re_{\rm M}^{\rm mph}$ for $M_{\star} \geq 10^{10}\ M_{\odot}$ galaxies, we cannot compare their merger rate with that from $N_{\rm c}^{\rm M}(z,10^{10}\ M_{\odot})$ because of the progenitor bias \citep{bell06,lotz08ff}. In morphological studies we are sensitive to the high distorted remnant phase of a major merger ($\mu \gtrsim 1/3$, \citealt{conselice06ff,lotz09t}), while in pair studies we see the pre-merger stage; that is, the mass of the future remnant is the sum of the two galaxies' masses in the pair. This also was noted by \citet{genel09}, who compare the merger rate per progenitor (i.e., close pairs) and descendant (i.e., morphological criteria) dark matter halo in the Millennium Simulation \citep{springel05}. They find that these two merger rates are different quantities, so close pairs and morphological studies do not measure the same merger rate. We therefore compare the morphological merger rate of $M_{\star} \geq 10^{10}\ M_{\odot}$ galaxies with that from $N_{\rm c}^{\rm M}$ obtained with $\log(M_{\star,1}^{\rm sel}/M_{\odot}) = 9.7$, $\log(M_{\star,2}^{\rm sel}/M_{\odot}) = 9.4$, and $\mu \geq 1/3$. With this definition, the merger of two galaxies with limiting masses is $\log(M_{\star}/M_{\odot})\sim 9.9$, a mass that could increase owing to the induced star formation in gas-rich mergers \citep[e.g.,][]{lin07,li08,knapen09,robaina09}. Finally, we apply a $T_{\rm pair} = 0.75$ Gyr delay to the close pairs data points to mimic the redshifts at which the highly distorted remnants of these close pairs systems could be observed. The higher redshift point at $z = 0.875$ becomes $z = 0.725$, while the point at $z = 0.4$ becomes $z = 0.335$. In summary, both morphological and close pair studies now provide the {\it rate of major merger remnants with $M_{\star} \geq 10^{10}\ M_{\odot}$} (i.e., per descendant galaxy), and we can compare each of them.

\begin{figure}[t!]
\resizebox{\hsize}{!}{\includegraphics{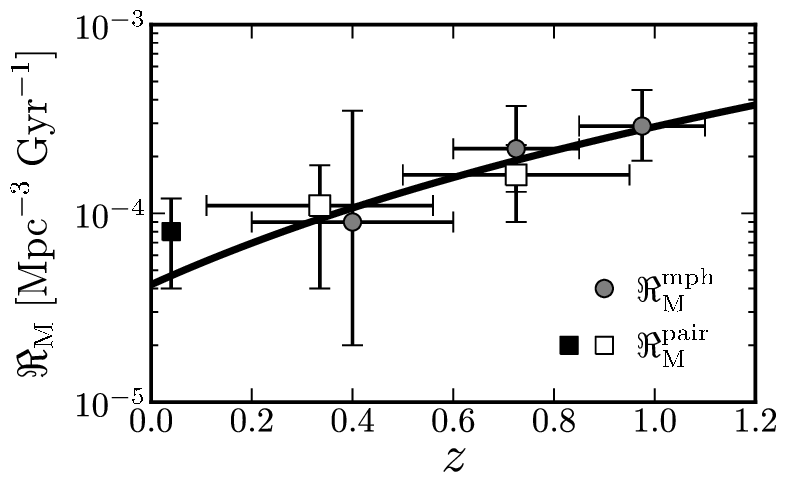}}
\caption{Major merger rate vs redshift for $M_{\star} \gtrsim 10^{10}\ M_{\odot}$ remnant galaxies. White squares are from the present work by pair statistics, grey circles are from L09 by morphological criteria, and the black square is from \citet{domingue09} by pair statistics. The black solid line is the best power-law fit to the $z > 0.2$ data and refers mainly to gas-rich major mergers (see text for details).}
\label{mrfig}
\end{figure}

We summarize the final values of $\Re_{\rm M}^{\rm pair}$ from this work and those of $\Re_{\rm M}^{\rm mph}$ from L09 in Table~\ref{mrtab}, and show them in Fig.~\ref{mrfig}. We see that both methodologies provide similar merger rates within the error bars in the range under study, as expected if we are observing two different phases of the same physical process in the same field (i.e., we minimize the field-to-field variance between both measurements). This reconciles the morphological merger rates with those from pair statistics and lends credibility to the asymmetry index $A$ as a major merger indicator.

We parametrize the merger rate evolution with a power-law function,
\begin{equation}
\Re_{\rm M} = \Re_{\rm M}(0)(1+z)^{n}\label{mreq}.
\end{equation}
The fit to all the data in Table~\ref{mrtab} yields 
\begin{equation}
\Re_{\rm M} = (0.42 \pm 0.07) (1+z)^{2.8 \pm 0.3} \times 10^{-4}\ {\rm Mpc}^{-3}\ {\rm Gyr}^{-1}.\label{mreqfit}
\end{equation}
These parameters are substantially different from those in L09, who find $\Re_{\rm M}(0) = (0.29 \pm 0.06) \times 10^{-4}$ Mpc$^{-3}$ Gyr$^{-1}$ and $n = 3.5 \pm 0.4$ from morphological data alone.

In the previous discussion we do not consider that the close pair methodology is sensitive to gas-rich (wet + mixed) and spheroidal-spheroidal (dry) mergers, while the morphological merger rate only refers to gas-rich major mergers (\citealt{conselice06ff,lotz09t}; L\'opez-sanjuan et al., in prep.). However, at $z \gtrsim 0.3$ the fraction of dry mergers with respect to the total is $\sim 10$\% (\citealt{lin08}, see also \citealt{deravel09}), increasing their importance at $z < 0.2$ and being $\sim 25$\% of the total at $z \sim 0.1$ \citep{lin08}. This implies that $\Re_{\rm M}^{\rm pair}$ must be similar to $\Re_{\rm M}^{\rm mph}$ in the range under study, but higher at low redshifts, when dry mergers become more important. We can compare the gas-rich merger rate at $z = 0.04$ from the previous fit, $\Re_{\rm M}(0.04) = 0.52 \times 10^{-4}$ Mpc$^{-3}$ Gyr$^{-1}$, with the {\it total} major merger rate inferred from the number of companions ($N_{\rm c} = 0.011 \pm 0.003$) observed by \citet{domingue09} at that redshift and stellar mass, $\Re_{\rm M} = (0.8 \pm 0.4) \times 10^{-4}$ Mpc$^{-3}$ Gyr$^{-1}$ (black square in Fig.~\ref{mrfig}). The latter is higher than the former and suggests that $\sim 35$\% of the mergers at $z = 0.04$ are dry, in agreement with the expected tendency (see \citealt{wen09} for an estimation of the dry merger rate of luminous early-type galaxies at $z \sim 0.1$).

\begin{table}
\caption{Major merger rate of $M_{\star} \gtrsim 10^{10}\ M_{\odot}$ galaxies in GOODS-S}
\label{mrtab}
\begin{center}
\begin{tabular}{lcc}
\hline\hline
Redshift & $\Re_{\rm M}^{\rm mph}$ & $\Re_{\rm M}^{\rm pair}$ \\
 & $(10^{-4}\ {\rm Mpc^{-3}}$ Gyr$^{-1})$ & $(10^{-4}\ {\rm Mpc^{-3}}$ Gyr$^{-1})$\\
\hline
$0.335$	& \ldots 		& $1.1\pm0.7$ \\
$0.4$	& $0.9^{+2.6}_{-0.7}$ 	& \ldots \\
$0.725$	& \ldots 		& $1.6\pm0.7$ \\
$0.725$	& $2.2^{+1.5}_{-0.9}$ 	& \ldots \\
$0.975$	& $2.9^{+1.6}_{-1.0}$ 	& \ldots \\
\hline
\end{tabular}
\end{center}
\end{table}

\section{The role of mergers in the evolution of intermediate-mass early types}\label{discussion}
In a previous work, \citet{clsj10megoods} have studied the number density evolution of early-type (ET, E/S0/Sa) galaxies of $\log(M_{\star}/M_{\odot}) \geq 10$ at $z \leq 1$, finding that these galaxies increase their comoving number density ($\rho_{\rm ET}$) by a factor of 5 from $z = 1$ to the present. Comparing the evolution of the early-type population between $z_{\rm sup}$ and $z_{\rm inf} < z_{\rm sup}$, named $\rho^{\rm new}_{\rm ET}(z_{\rm inf},z_{\rm sup}) = \rho_{\rm ET}(z_{\rm inf}) - \rho_{\rm ET}(z_{\rm sup})$, against the gas-rich major merger rate in the same redshift range, we can define $f_{\rm ET,M}$, the fraction of new early types that appear between $z_{\rm sup}$ and $z_{\rm inf}$ because of gas-rich major mergers:
\begin{equation}
f_{\rm ET,M}(z_{\rm inf},z_{\rm sup}) = \frac{\rho_{\rm rem}^{\rm M}(z_{\rm inf},z_{\rm sup})}{\rho_{\rm ET}(z_{\rm inf}) - \rho_{\rm
ET}(z_{\rm sup})},\label{fremetm}
\end{equation}
where $\rho_{\rm rem}^{\rm M}(z_{\rm inf},z_{\rm sup})$ is the number density of gas-rich major merger remnants,
\begin{equation}
\rho_{\rm rem}^{\rm M}(z_{\rm inf},z_{\rm sup}) = \int_{z_{\rm inf}}^{z_{\rm sup}}\Re_{\rm M}(0)(1+z)^{n-1}\frac{{\rm d}z}{H_0 E(z)},
\end{equation}
where $E(z) = \sqrt{\Omega_{\Lambda} + \Omega_{M}(1+z)^3}$ in a flat universe, and $\Re_{\rm M}(0)$ and $n$ are the merger rate parameters in Eq.~(\ref{mreq}). We assume that gas-rich merger remnants are early-type galaxies \citep{naab06ss,rothberg06a,rothberg06b,hopkins08ss,hopkins09disk}.

Using the L09 gas-rich major merger rate, \citet{clsj10megoods} infer that $f_{\rm ET,M}(0,1) = 17^{+10}_{-7}$\% for early-type galaxies of $\log(M_{\star}/M_{\odot}) \geq 10$. As we have shown in Sect.~\ref{mrpair}, when we join the major merger rate by close pairs statistics from this work with the one in L09, derived from morphological criteria, the merger rate parameters change substantially from those derived by L09 using morphological information alone. However, the number density of major merger remnants remains similar in the two determinations, varying from $\rho_{\rm rem}^{\rm M}(0,1) = (8.5^{+3.8}_{-2.8}) \times 10^{-4}$ Mpc$^{-3}$ with L09 parameters to $\rho_{\rm rem}^{\rm M}(0,1) = (9.1^{+2.9}_{-2.4}) \times 10^{-4}$ Mpc$^{-3}$ with the parameters from the present paper (Eq.~[\ref{mreqfit}]), which implies $f_{\rm ET,M}(0,1) = 18^{+11}_{-7}$\%. Both morphologies \citep{clsj10megoods} and close pairs (this work) therefore agree in yielding consistently low values for the fraction of early types that appear between redshifts 1 and 0 due to major mergers.

\subsection{Estimating the minor merger rate}\label{mmrm}
Since gas-rich major mergers cannot explain the rise in $M_{\star} \geq 10^{10}\ M_{\odot}$ early-type galaxies since $z \sim 1$, \citet{clsj10megoods} suggest minor mergers and secular processes (e.g., bars, disc instabilities, gas exhaustion, or morphological quenching) as the main path in that evolution (see also \citealt{bundy09red,oesch09}). In Sect.~\ref{ncmass} we find that the number of minor companions ($N_{\rm c}^{\rm m}$) is roughly twice the number of major companions ($N_{\rm c}^{\rm M}$). This implies that $N_{\rm c}(1/10 \leq \mu < 1/3) = N_{\rm c}^{\rm m} - N_{\rm c}^{\rm M} \sim N_{\rm c}^{\rm M}$, so we can estimate the minor merger rate ($\Re_{\rm mm}$), defined as the merger rate of galaxies with $1/10 \leq \mu < 1/3$. We denote $\Re_{\rm m} = \Re_{\rm M} + \Re_{\rm mm} = R \times \Re_{\rm M}$ as the total (major + minor) merger rate.

We apply Eq.~(\ref{mrpar}) to obtain $\Re_{\rm mm}$. We assume that parameters for minor companions are similar to those for major companions, except for the merger time scale $T_{\rm pair}$. From N-body hydrodynamical simulations, \citet{lotz09t} find $T_{\rm pair}^{\rm mm} \sim 1.5 \times T_{\rm pair}^{\rm M}$ for $5h^{-1}$ kpc $< r_{\rm p} < 20h^{-1}$ kpc close pairs. With the previous assumptions we infer that $\Re_{\rm mm} \sim 0.7 \times \Re_{\rm M}$, and $\Re_{\rm m} \sim 1.7 \times \Re_{\rm M}$. The latest cosmological models \citep[e.g.,][]{stewart08,gongar09,hopkins09fusbul} predict $R \sim 2$, similar to our estimation $R \sim 1.7$. However, simulations refer to minor mergers selected by {\it baryonic} (gas + stellar) mass, while our estimation is from minor mergers selected by {\it stellar} mass. Although baryonic masses are more uncertain than stellar ones, it is interesting to compare the baryonic and the stellar merger rates, which could be quite different \citep{stewart09}. Moreover, the baryonic merger rate can be compared to the fraction of galaxies with distorted kinematics at $z \sim 0.6$ \citep{neichel08}, which are described well by merger simulations ($\mu_{\rm baryonic} \gtrsim 1/5$, \citealt{hammer09}).

\subsection{Mergers vs secular processes in the evolution of early-type galaxies}
Following the Sect.~\ref{discussion} steps with the combined major and minor merger rates estimated in the previous section, we infer that $\sim30$\%, and up to $\sim$50\%, of the early-type (E/S0/Sa) galaxies of $M_{\star} \geq 10^{10}\ M_{\odot}$ that appear since $z \sim 1$ may have undergone a major or minor merger event. This is an upper limit because a single minor merger does not transform a late-type galaxy into an early type \citep{hopkins09disk}, but increases the S\'ersic index of the galaxy \citep{eliche06,bournaud07}. This result suggests that the other $\sim 50$\% of the new early types appear due to secular processes (see also \citealt{bundy09red}). 

That a large fraction of S0-Sb discs host pseudo-bulges \citep{kormendy04} also points to secular processes for the growth of the red sequence at $z < 1$.  And the strong similarity of disc and nuclear colours in disc galaxies up to $z\sim 0.8$ \citep{palmero08} likewise argues for a fading process that does not destroy the disc.  The process leading to such an evolution may be suggested by the mentioned colour similarity of (the inner parts of the) disc and nucleus/bulge.  Both star formation and the subsequent fading must be to some degree coordinated throughout the galaxy, so the quenching of star formation may simply be due to gas exhaustion \citep{zheng07,bauer09}.  Interestingly, the star formation in the galaxy, leading to the growth of the central bulge component, may lead to this fading as the growth of the central potential stabilizes the gas component preventing disc fragmentation \citep[morphological quenching,][]{martig09}.  Disc fragmentation contributes both to star formation and to the growth of the central bulge, as shown by the evolution of chain and tadpole galaxies \citep{elmegreen08,ceverino09,bournaud09}.  
 
These results refer to intermediate-mass galaxies ($M_{\star} \sim 4\times10^{10}\ M_{\odot}$) at $z \lesssim 1$, but the picture is different at higher stellar masses. On the one hand, $M_{\star} \gtrsim 10^{11}\ M_{\odot}$ galaxies have higher pair fractions than less
massive ones \citep{deravel09,bundy09}, and red pairs are more common at these masses \citep{bundy09}. This suggests that dry mergers are an important process in the evolution of massive
galaxies since $z \sim 1$ \citep[e.g.,][]{bell04,lin08,ilbert10}. On the other hand, the size
\citep{trujillo07,buitrago08,vanderwel08esize,vandokkum10} and velocity
dispersion evolution \citep{cenarro09} of $M_{\star} \gtrsim 10^{11}\ M_{\odot}$
early-type galaxies since $z \sim 2$ also supports the importance of mergers, especially
the impact of minor mergers on the evolution of these systems
\citep{bezanson09,naab09,hopkins09size}. In addition, residual star formation in early-type (E/S0) galaxies at $z \lesssim 0.7$ can also be explained by minor merging \citep{kaviraj09,kaviraj10}. This problem has also been analysed by \citet{eliche10I}, who have modelled the evolution of luminosity function backwards in time for $M_{\star} \gtrsim 10^{11}\ M_{\odot}$ galaxies, selected according to their colours (red/blue/total) and their morphologies.  They find that the observed luminosity function evolution can be explained naturally by the observed gas-rich and dry major merger rates and that 50-60\% of
today's E/S0 in this mass range were formed by major mergers at $0.8 < z < 1$,
with a small number evolution since $z = 0.8$ (see also \citealt{cristobal09, vanderwel09, ilbert10,eliche10II}). The gas-rich major merger fractions assumed by \citet{eliche10I} are those
from L09 for $B$-band selected galaxies ($M_B \leq -20$), which were obtained in
a similar way than the morphological merger fractions used through present paper.

This makes $M_{\star}^{*} \sim 10^{11}\ M_{\odot}$ \citep{pgon08} a transition mass at $z \lesssim 1$: at higher masses mergers are an important process in the evolution of early-type galaxies, while other mechanisms dominate the observed evolution at lower masses.

\section{Conclusions}\label{conclusion}
In this paper we have developed a new method, which is based on the one used widely over spectroscopic surveys, to determine the mean number of companions per galaxy ($N_{\rm c}$) over current spectro-photometric surveys. We tested our method in a local, volume-limited sample from MGC spectroscopic catalogue. We find that the method provides reliable $N_{\rm c}$ values when either photometric redshift errors are smaller than $\sigma_{z_{\rm phot}}^{\rm max}$ or the spectroscopic completeness of the sample is higher than $f_{\rm spec}^{\rm min}$, with these limits depending on the sample under study. For typical SDSS samples with $\sigma_{z_{\rm phot}} = 0.02$, we find $f_{\rm spec}^{\rm min} \sim 0.2$, while we find $f_{\rm spec}^{\rm min} \sim 0.4$ for our GOODS-S catalogue.

We studied the number of companions in $B$-band luminosity-selected samples, finding that

\begin{itemize}
\item $N_{\rm c}$ depends on search radius as expected from correlation function, $N_{\rm c}\propto r_{\rm p}^{1.3}$, and their values are similar to those expected from a spectroscopic sample.

\item The values of $N_{\rm c}$ for different luminosity selections are in excellent agreement with those in the literature when the same selection criteria and pair definition are applied. We find that the number of companions decreases when luminosity increase and that the $N_{\rm c}$ of $M_B \leq -20$ galaxies evolves faster ($m = 2.5$) than for $M_B \leq -19$ galaxies ($m = 1.6$). In addition, this evolution becomes slower, $m = 0.7$, when luminosity evolution is taken into account.
\end{itemize}

We applied our new methodology to estimate the relation between the close major companions ($N_{\rm c}^{\rm M}$, mass ratio $\mu \geq 1/3$) and the close minor companions ($N_{\rm c}^{\rm m}$, mass ratio $\mu \geq 1/10$) of $M_{\star} \geq 10^{10}\ M_{\odot}$ galaxies. We find that $N_{\rm c}^{\rm m}$ is roughly two times $N_{\rm c}^{\rm M}$. Studies in more extended fields are needed to better understand the minor-to-major ratio and their possible evolution with redshift.

Finally, we compared the merger rates derived by close pairs with those by morphological criteria for $M_{\star} \geq 10^{10}\ M_{\odot}$ galaxies, finding that both are similar when the progenitor bias is taken into account. We estimate that the total (major + minor) merger rate is $\sim$1.7 times the major merger rate. After comparing the merger rate with the observed structural evolution of GOODS-S galaxies, we infer that up to $\sim50$\% of the new early-type galaxies appeared since $z \sim 1$ could have undergone a merger event (major or minor), and we refer to star formation fading in the disc of spiral galaxies to explain the other $\sim50$\%.

Our new methodology can be applied to extensive spectro-photometric surveys such as AEGIS, COSMOS, or VVDS, which have an order of magnitude more galaxies than the present catalogue. This will allow better determination of the number of major and minor companions and their dependence on redshift and other galactic properties, such as luminosity, mass, colour, or environment.

\begin{acknowledgements}
We dedicate this paper to the memory of our six IAC colleagues and friends who
met with a fatal accident in Piedra de los Cochinos, Tenerife, in February 2007.
With a special thanks to Maurizio Panniello, whose teachings of \texttt{python}
were so important for this paper. 

We thank the anonymous referee for his/her pertinent comments that have improved the original manuscript. 
We also thank Ignacio Trujillo, Carmen Eliche-Moral, and Rub\'en Sanchez-Janssen for 
useful discussions and suggestions. 

This work was supported by the Spanish Programa Nacional de Astronom\'\i a y
Astrof\'{\i}sica through project number AYA2006--12955, AYA2006--02358, and AYA
2006--15698-C02-02. This work was partially funded by the Spanish MEC under the
Consolider-Ingenio 2010 Programme grant CSD2006-00070: First Science with the GTC
(http://www.iac.es/consolider-ingenio-gtc/).

This work is based in part on observations made with the {\it Spitzer} Space
Telescope, which is operated by the Jet Propulsion Laboratory, Caltech under
NASA contract 1407.

This work uses the Millennium Galaxy Catalogue, 
which consists of imaging data from the
Isaac Newton Telescope and spectroscopic data from the Anglo
Australian Telescope, the ANU 2.3m, the ESO New Technology Telescope,
the Telescopio Nazionale Galileo, and the Gemini North Telescope. The
survey was supported through grants from the Particle Physics and
Astronomy Research Council (UK) and the Australian Research Council
(AUS). The data and data products are publicly available from
http://www.eso.org/~jliske/mgc/ or on request from J. Liske or
S.P. Driver.

P. G. P. G. acknowledges support from the Ram\'on y Cajal Programme financed by
the Spanish Government and the European Union.
\end{acknowledgements}

\bibliography{biblio}
\bibliographystyle{aa}
\end{document}